\documentclass[11pt,a4paper]{deepseek}
\newcommand{\ReportTitle}{MuLaCover: Flexible Cover Song Generation via Symbolic Tonal Control}
\newcommand{\ReportPDFTitle}{MuLaCover: Flexible Cover Song Generation via Symbolic Tonal Control}

\newcommand{\ReportPDFAuthor}{MuLa Labs, Vera Praxis Lab}
\newcommand{\ReportInstitution}{MuLa Labs, Vera Praxis Lab}
\newcommand{\ReportEmail}{contact@mulalabs.ai}
\newcommand{\ReportGitHub}{https://github.com/HeartMuLa/MuLaCover}

\newif\ifReportLogo
\ReportLogotrue
\newcommand{\ReportLogoFile}{vera-praxis-logo.pdf}

\newif\ifReportContents
\ReportContentstrue

\usepackage{bm,mathtools}
\usepackage{array,multirow}
\usepackage{float}
\usepackage[bottom]{footmisc}
\usepackage{CJKutf8}
\usepackage{tikz}
\usetikzlibrary{arrows.meta,backgrounds,fit,positioning,shapes.geometric}
\usepackage[most]{tcolorbox}
\usepackage[numbers,compress]{natbib}
\usepackage{setspace}

\definecolor{MethodBlue}{HTML}{175A86}
\definecolor{MethodDark}{HTML}{17324D}
\definecolor{MethodGray}{HTML}{66717B}
\definecolor{MethodLight}{HTML}{EEF5F8}
\definecolor{MethodRule}{HTML}{BCD2DE}
\definecolor{KeepGreen}{HTML}{24785A}
\definecolor{KeepLight}{HTML}{EDF7F2}
\definecolor{DropRed}{HTML}{A94343}
\definecolor{DropLight}{HTML}{FBEFEF}
\definecolor{AuditSlate}{RGB}{51,65,85}
\definecolor{AuditPaper}{RGB}{248,250,252}
\definecolor{AuditRule}{RGB}{203,213,225}

\newtcolorbox{promptbox}[1]{  enhanced,
  breakable,
  colback=AuditPaper,
  colframe=AuditRule,
  boxrule=0.5pt,
  arc=1.5mm,
  boxsep=0pt,
  left=9pt,
  right=9pt,
  top=7pt,
  bottom=7pt,
  title={#1},
  fonttitle=\sffamily\bfseries\small,
  coltitle=white,
  colbacktitle=AuditSlate,
  toptitle=5pt,
  bottomtitle=5pt,
  lefttitle=9pt,
  fontupper=\footnotesize,
  before skip=12pt,
  after skip=12pt,
  bottomrule at break=0pt,
  toprule at break=0pt,
  before upper={\parindent=0pt \parskip=2pt},
}

\newtcolorbox{jsonbox}{  colback=white,
  colframe=AuditRule,
  boxrule=0.4pt,
  arc=1mm,
  left=7pt,
  right=7pt,
  top=5pt,
  bottom=5pt,
  before skip=7pt,
  after skip=7pt,
}

\hypersetup{
  colorlinks=true,
  linkcolor=blue,
  citecolor=blue,
  urlcolor=blue,
  hypertexnames=false
}

\setlist[itemize]{leftmargin=*,topsep=3pt,itemsep=2pt}
\graphicspath{{figures/}}
\newcommand{\ReportHeader}{%
  \ifReportLogo
    \raisebox{-0.24\height}{\includegraphics[width=72pt]{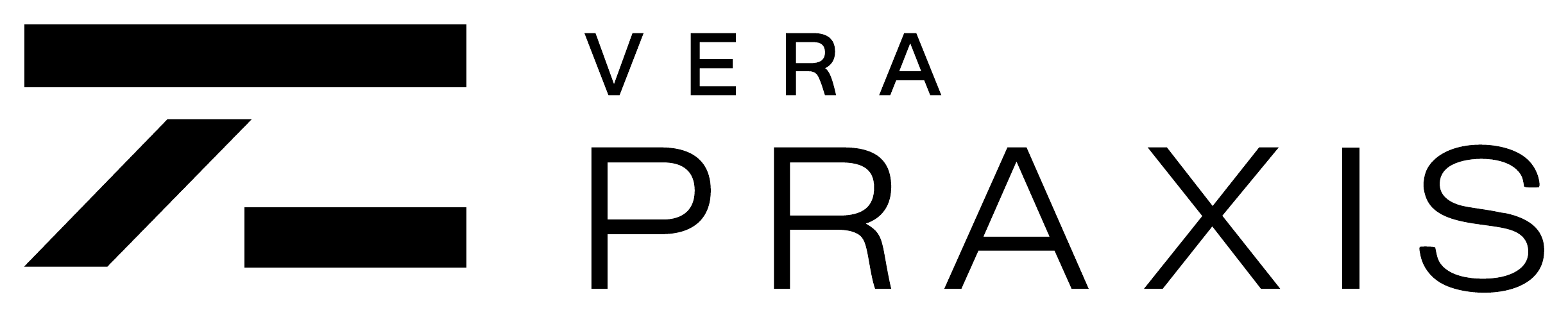}}%
  \else
    {\sffamily\bfseries\small\ReportInstitution}%
  \fi}
\fancypagestyle{firststyle}{%
  \fancyhf{}
  \fancyhead[L]{\ReportHeader}
  
  }
\fancypagestyle{report}{%
  \fancyhf{}
  \fancyhead[L]{\ReportHeader}
  \fancyfoot[C]{\footerfont\thepage}
  
  }
\title{\centering\ReportTitle}
\author{\centering\normalfont
  \ReportInstitution\par\vspace{7pt}
  {\small\href{mailto:\ReportEmail}{\texttt{\ReportEmail}}}\par\vspace{4pt}
  {\small\href{\ReportGitHub}{\texttt{\ReportGitHub}}}\par\vspace{4pt}
  {\small\href{https://verapraxis.ai/mulalabs}{\texttt{https://verapraxis.ai/mulalabs}}}\par}
\date{}
\hypersetup{pdftitle={\ReportPDFTitle},pdfauthor={\ReportPDFAuthor}}


\usepackage{placeins}

\usepackage{nicefrac}
\usetikzlibrary{calc}

\newcommand{\ours}{MuLaCover}

\usepackage{etoolbox}

\begin{abstract}

A cover song re-renders an existing piece, most commonly by preserving its tonal content—melody and chord progression—while reshaping other musical attributes. This suggests a natural formulation for generative cover synthesis with pretrained music language models: explicitly control tonal content, while specifying lyrics and style through text. Existing approaches condition on frame-level features, tightly coupling generation to the reference’s timing and structure, and limiting flexibility such as tempo variation, structural rearrangement, or partial conditioning. To this end, we present \ours{}, a cover generation model that conditions a pretrained text-to-song foundation model on an editable symbolic lead sheet. Our design enables alignment-free control: generated outputs preserve the tonal signature of the source without frame-level alignment, allowing flexible timing, structure, and segment-level generation. We further introduce a training curriculum with partially mismatched audio–symbolic pairs, improving diversity and robustness. We evaluate \ours{} with objective metrics, standard subjective ratings, and in-depth expert interviews that provide fine-grained diagnostic insights. Experiments show that \ours{} achieves state-of-the-art performance among open-source systems and is competitive with leading commercial models such as Suno-v5.5. Audio samples and codes are available at our project page \url{https://mulalabs.ai/mulacover/}.

\end{abstract}

\begin{document}
\makeatletter
\begingroup
\setlength{\parindent}{0pt}
\setlength{\parskip}{0pt}
\begin{center}
{\fontsize{19}{23}\selectfont\bfseries MuLaCover: Flexible Cover Song Generation\\[1pt]
via Symbolic Lead Sheet Control\par}
\vspace{9pt}
{\fontsize{11}{14}\selectfont\ReportInstitution\par}
\vspace{5pt}
{\fontsize{9}{11}\selectfont
\href{mailto:\ReportEmail}{\ReportEmail}\enspace\textcolor{gray}{|}\enspace
\href{\ReportGitHub}{GitHub}\enspace\textcolor{gray}{|}\enspace
\href{https://verapraxis.ai/mulalabs}{MuLa Labs}\par}
\end{center}
\vspace{3pt}
{\fontsize{10.5}{13}\selectfont\bfseries Abstract\par}
\vspace{4pt}
{\fontsize{11}{14}\selectfont\theabstract\par}
\endgroup
\thispagestyle{firststyle}
\makeatother
\pagestyle{report}

\begin{figure}[H]
  \captionsetup{font=footnotesize,skip=7pt}
  \centering
  \includegraphics[width=\textwidth]{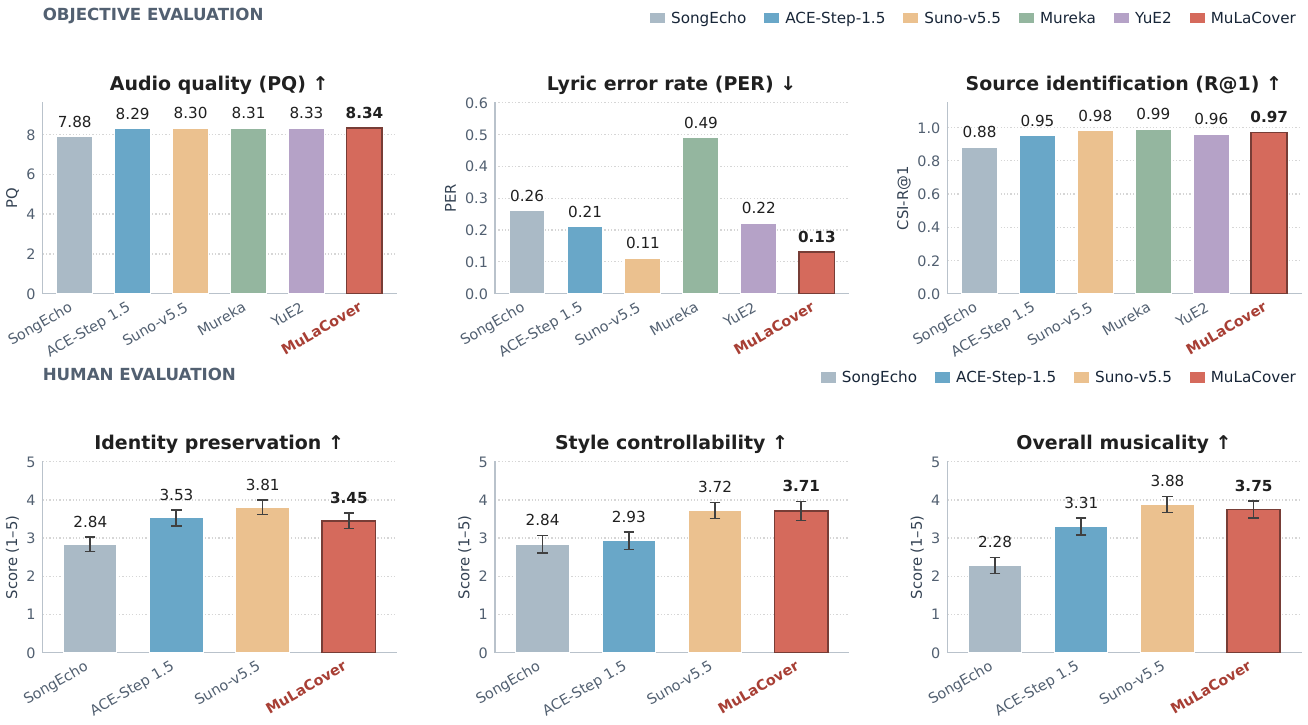}
  \caption{Full-song cover generation: objective results for six systems (top; Mureka: 248 outputs, others: 256) and listening scores for four systems (bottom; mean $\pm$ 95\% CI, 22 participants). Mureka and YuE2 were not included in the listening study. Data: Tables~\ref{tab:obj_main} and~\ref{tab:subj_main}.}
  \label{fig:results_overview}
\end{figure}

\ifReportContents
  \clearpage
  \phantomsection
  \pdfbookmark[0]{\contentsname}{report-contents}
  \begingroup
    \hypersetup{linkcolor=red,linktoc=section}
    \setlength{\parskip}{0pt}
    \begin{spacing}{0.95}
      \tableofcontents
    \end{spacing}
  \endgroup
  \clearpage
\fi

\section{Introduction}

A \emph{cover song} is a re-rendering of an existing piece that preserves
its musical identity while freely reshaping every other aspect of the
recording. Decades of
cover-song-identification
research~\cite{csi_survey,chroma_features_beat_tracking,serra_chroma_binary,CNN_CSI,bytecover,bytecover3,coverhunter} converge on a
clean characterization of this identity: across renditions, key, tempo,
instrumentation, vocal timbre, song structure, and even lyrics may all
change; what typically stays invariant is the \emph{tonal content} --- the melody
and the underlying chord progression. This invariant/variant taxonomy
suggests a natural recipe for cover \emph{generation}: condition a
pretrained text-to-song foundation
model~\cite{yuan2025yue,gong2025ace,yang2026heartmula} on the tonal
content of a source rendition, and let its lyrics and style prompts
shape everything else.

The unresolved design question is how to encode the tonal-content
signal. Existing audio-domain cover systems answer it with frame-level
features extracted from the source recording itself, such as vocal $F_0$
contours~\cite{lisongecho} or quantized audio
latents~\cite{gong2026ace}, sampled at the same rate as the target
audio. This choice imposes a strict \emph{frame-wise alignment} between source
and cover: the cover cannot run at a different tempo, cannot rearrange
structure, and cannot draw on a partial reference such as a
single verse or chorus.

We argue that the right tonal-content representation is the
\emph{symbolic lead sheet} long used in music information retrieval \cite{bai2024seed,ls1,ls2}: a
melody line paired with a chord progression, which by construction
abstracts away surface acoustics and retains only the typically cover-invariant
musical skeleton. Crucially, before feeding the lead sheet into the
model, we time-stretch it so that all sequences share a canonical
tempo. This BPM normalization is what frees the conditioning signal
from any frame-wise tie to the source recording: it leaves the cover
model free to set its own pace, and, via attention masking over the
symbolic sequence, lets the model accept any sub-segment of the song as
a partial reference. We inject this normalized symbolic signal into a
pretrained text-to-song backbone through a gated cross-attention
module, yielding our system \textbf{\ours{}}.

Two issues remain after the architectural design. The first concerns data.
Most text-to-song models~\cite{yuan2025yue,ning2025diffrhythm,leilevo,gong2025ace,yang2026heartmula,lisongecho} are trained on recordings whose
lyrics, style, and audio necessarily come from the same single
rendition; the model has never been asked to render the tonal content
of one rendition under the lyrics and style of a \emph{different}
one, yet precisely this mismatch is what cover generation demands.
We close this supervision gap through \emph{cross-cover finetuning} on
a curated paired dataset that pairs the tonal content of one rendition
with the lyrics, style, and audio of another. The second concerns evaluation.
Cover quality is multi-faceted and idiomatic, but
an aggregate Mean Opinion Score (MOS) listening study reveals \emph{whether} a system is good
without explaining \emph{why}, and offers little guidance for system
iteration. We therefore complement listening-study questionnaires with a structured
\emph{expert-interview} protocol that elicits diagnostic, comparative
judgments from trained listeners. Together with objective metrics, these subjective signals yield a comprehensive evaluation.

We evaluate \ours{} on two cover tasks. On the standard
\emph{full-song cover} task, \ours{} is state-of-the-art among
open-source systems on objective metrics of musical quality, lyric and
style controllability, and cover identifiability, and is broadly on par
with the leading closed-source commercial system Suno-v5.5; the
listening study and expert interviews corroborate these findings. As a
natural consequence of its alignment-free conditioning, \ours{} further
tackles \emph{partial-reference cover generation}, in which only a
fragment of the reference is provided and the model produces a full-song
cover that preserves the fragment's tonal content while freely
re-creating the rest, a task on which \ours{} achieves superior performance over existing open-sourced baselines across three fragment
configurations in objective evaluation.

Our main contributions are:
\begin{itemize}
    \item We formulate cover generation as conditioning a pretrained
          text-to-song foundation model on the BPM-normalized symbolic lead
          sheet of a source rendition, and realize it as \ours{}, which injects
          the signal through a gated cross-attention module. The design removes
          the strict frame-wise alignment that has bound prior cover systems to
          the source's exact timing and structure, and supports partial-reference
          cover generation as a natural by-product.

    \item We identify cross-version pairing of content and style as the
          missing supervision signal for cover generation, and provide a curated
          paired dataset together with a cross-cover finetuning curriculum that
          turns a generic text-to-song model into a cover generator.

    \item We evaluate \ours{} with objective metrics, listening-study
          questionnaires, and a structured expert-interview protocol that
          complements aggregate scores with fine-grained, comparative
          diagnostics. The combined evaluation establishes \ours{} as
          state-of-the-art among open-source cover systems and broadly on par
          with the closed-source commercial system Suno-v5.5.
\end{itemize}

\section{Related Work}
\label{sec:related}

\subsection{Cover Song Identification}

Cover song identification (CSI) is the music information retrieval task of retrieving
cover versions of a query song from a database \cite{csi_survey}. Across versions, key,
tempo, structure, instrumentation, timbre, and lyrics may all change \cite{csi_survey,chroma_features_beat_tracking,serra_chroma_binary,CNN_CSI,bytecover,bytecover3,coverhunter}; the
most stable cross-version identifier is the \emph{tonal content} \cite{serra_chroma_binary}, namely the
melody \cite{tsai2005query_melody} and the underlying chord progression or chroma features
\cite{izmirli2005tonal,chroma_features_beat_tracking}, which serves as the primary
cross-version identifier. This invariant-vs-variant decomposition has shaped algorithmic
design throughout the CSI literature. Classical rule-based systems
\cite{chroma_features_beat_tracking,serra_chroma_binary} extract tonality-oriented
descriptors such as chroma features \cite{feat_chroma} or HPCP sequences \cite{feat_hpcp},
and pair them with similarity computations that explicitly account for the varying
attributes: transposition handles key shifts, and dynamic time warping handles tempo
differences \cite{serra_chroma_binary}.
Modern deep CSI systems \cite{CNN_CSI,remove,bytecover,bytecover3,coverhunter} preserve
the same inductive bias but realise it through learned representations, training networks
on contrastive or classification objectives with pooling, matching, and augmentation
strategies (e.g.\ frequency-axis rolling for pitch shifts, speed perturbation, and
spectrogram masking) deliberately designed so that embeddings remain invariant to key,
tempo, and structural variations between versions. Viewed through a generative lens,
the variant/invariant taxonomy of CSI is exactly a \emph{content/style} decomposition of
a song: tonal content is the content, and everything else (instrumentation, timbre, tempo,
structure, lyrics) is the style that a cover artist typically reshapes, although where
this line is drawn varies from cover to cover. Cover song generation
is then the \emph{style transfer} counterpart of CSI, and as such belongs to the broader
space of controllable music generation reviewed next.

\subsection{Controllable Music Generation}

In the audio domain, controllable music generation has advanced rapidly through
language-model and diffusion systems \cite{agostinelli2023musiclm,copet2023simple_musicgen,ning2025diffrhythm,gong2025ace,lei2024songcreator,yuan2025yue,leilevo,yang2026heartmula}.
Control here is predominantly textual---style descriptions and lyric prompts---with
finer-grained content signals such as melody and chord progression only recently
emerging as auxiliary conditions
\cite{lin2023content_cocomulla,tsaimusecontrollite,lan2024musicongen,lisongecho}.
Symbolic music generation, by contrast, has long offered interpretable, disentangled
control at global (time signature, tempo, instrumentation \cite{Musecoco,mmt,figaro}),
local (chord progression, compositional texture \cite{polydis,polyffusion,yi2026vitex}),
and hierarchical \cite{whole_song} scales. For cover song generation, symbolic piano cover systems
\cite{wang2022audio2midi,zhaobossa,tan2024picogen} remain narrow in scope. Audio-domain
systems such as SongEcho \cite{lisongecho} and ACE-Step~1.5 \cite{gong2026ace} condition
on a temporal feature of the source sampled at the same rate as the target audio features, enforcing strict
frame-wise alignment and leaving no degree of freedom in tempo or structure. \ours{}
removes this rigidity through flexible symbolic lead-sheet conditioning.

Since music composition is inherently a cascaded process, recent works chain multiple
specialised modules, each handling one sub-task (e.g.\ melody writing, harmonisation,
structural planning, singing synthesis, or sample retrieval) and passing intermediate
symbolic or structural representations to the next \cite{midisag,tomi}. While such
pipelines expose interpretable and editable intermediate stages, errors accumulate
across modules and the final audio is bounded by the individual components that
produce it. As a lead-sheet-conditioned audio generator, \ours{} can be intergrated naturally into such
pipelines to better support human--AI music co-creation.

\section{Method}
\label{sec:method}

\begin{figure}[t]
    \centering
    \includegraphics[width=\linewidth]{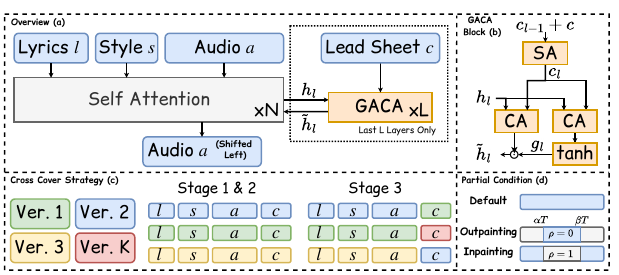}
    \caption{Overview of \ours{}. \textbf{(a)}~The pretrained text-to-song backbone has $N$ transformer layers consuming lyrics $\ell$, style $s$, and audio tokens $a$; at its last $L$ layers, the Gated Adaptive Cross-Attention (GACA) module injects the symbolic tonal-content condition $c$ into the hidden states $h_l$. \textbf{(b)}~Inside a GACA block, $c$ is refined by a layer-specific self-attention (SA) into $c_l$, then merged into $h_l$ through a cross-attention (CA) update modulated by a learned, $\tanh$-bounded gate $g_l$. \textbf{(c)}~Cross-cover strategy in Stage~3: colors denote different renditions of the same song. Stages~1--2 draw all four signals $\ell, s, a, c$ from the same rendition; Stage~3 draws $c$ from a different rendition than $\ell, s, a$. \textbf{(d)}~Partial conditioning via attention masking on $c$: $\rho{=}0$ keeps the interval $[\alpha T, \beta T]$ (\emph{outpainting}), $\rho{=}1$ keeps its complement (\emph{inpainting}); the default $(\alpha,\beta,\rho){=}(0,1,0)$ recovers the full-song setting.}
    \label{fig:reprise}
\end{figure}

\ours{} equips a decoder-only text-to-song backbone with tonal-content control.
Mainstream text-to-song foundation models \cite{yuan2025yue,yang2026heartmula}
already expose two controls --- a lyrics prompt $\ell$ along with a style prompt $s$ that conveys
non-lyric attributes such as genre and instrumentation --- and
autoregressively model the audio token sequence $a$ as their generation target. \ours{} adds
the tonal-content signal $c$ on top of these existing controls; an overview of the resulting architecture is shown in Figure~\ref{fig:reprise}(a).

Abstracting away the specific text-to-song instantiation, the remaining design problem
becomes one of sequence-to-sequence conditional injection. The backbone produces, at each
of its $N$ transformer layers, a sequence of hidden states $h_l \in \mathbb{R}^{S \times
        D}$ ($l = 1, \dots, N$, with $S$ the sequence length and $D$ the feature dimension); the new
control is itself a sequence $c \in \mathbb{R}^{T \times D}$ of length $T$. \ours{}
merges information from $c$ into $h_l$ at the last $L$ layers via a gated adaptive cross-attention module, leaving the
remaining $N{-}L$ layers and the original training loss intact.

Three components, each ablated independently in Section \ref{sec:exp_ablation}, complete the
system: (i)~a symbolic lead-sheet representation that defines the tonal-content sequence
$c$ (Section \ref{sec:cond_signal});
(ii)~the gated adaptive cross-attention injection module
(Section \ref{sec:injection}); and
(iii)~a three-stage curriculum that aligns the newly introduced parameters with the backbone
and exposes the model to cross-version training pairs in which the tonal-content signal
and the existing controls are drawn from different renditions of the same song
(Section \ref{sec:training}).

\subsection{Symbolic Lead Sheet Representation}
\label{sec:cond_signal}

A lead sheet, comprising a melody line and a chord progression, captures the tonal content
essential to the cover task and has been used in music generation \cite{bai2024seed,zhaobossa,midisag};
we adopt it as our conditioning signal $c$.

Given a reference recording, we extract the vocal melody and chord progression with
pretrained transcription models \cite{chang2024yourmt3+,jiang2019large_chord}. To prevent
the lead sheet from inheriting the source's specific tempo, and thereby to avoid imposing
any frame-wise alignment between $c$ and the generated audio, we estimate the source's
beats-per-minute (BPM) and time-stretch the lead sheet so that all sequences are normalized
to a canonical $\mathrm{BPM}{=}120$. We further quantize the symbolic timeline
at the resolution of a sixteenth note and denote the resulting sequence length as $T$.
This BPM normalization is essential to the system: removing it forces a strict frame-wise
correspondence between $c$ and the audio that exposes a sharp content--style tradeoff,
as we quantify in Section~\ref{sec:exp_ablation} (w/o BPM-norm).

After BPM normalization, the melody is encoded as a binary pianoroll
$M \in \{0,1\}^{T \times 256}$, whose $256 = 128 \times 2$ feature dimensions capture the
onset and sustain status of each of the $128$ MIDI pitches at every step; the chord
progression is encoded as a binary chromaroll $H \in \{0,1\}^{T \times 12}$ over the twelve
pitch classes. The two representations are projected into a shared embedding space and
concatenated along the feature dimension to form the conditioning sequence consumed
downstream:
\begin{equation}
    c \;=\; \bigl[\, M\, W_M \;\Vert\; H\, W_H \,\bigr] \;\in\; \mathbb{R}^{T \times D},
    \label{eq:condition}
\end{equation}
where $W_M \in \mathbb{R}^{256 \times D/2}$ and $W_H \in \mathbb{R}^{12 \times D/2}$ are
learned linear projections, $\Vert$ denotes feature-wise concatenation, and $D$ matches the
feature dimension of the backbone. Conditioning on the melody alone, without the
chord progression, is ablated in Section~\ref{sec:exp_ablation}
(w/o Chord); we also compare the symbolic lead sheet against the
frame-wise $F_0$ contour used by prior cover systems
\cite{lisongecho} in the same section
(w/ $F_0$).

\subsection{Gated Adaptive Cross-Attention Injection}
\label{sec:injection}

\ours{} injects the symbolic condition $c$ only into the last $L$ layers of the backbone
--- layers $N{-}L{+}1$ through $N$. The intuition is that adding tonal-content control
should only need to reshape the predicted audio-token distribution at the final stages of
generation; the lower $N{-}L$ layers, which carry the pretrained backbone's general
acoustic representations, can therefore be left structurally unchanged. The internal
structure of the resulting GACA block is illustrated in Figure~\ref{fig:reprise}(b).
We compare this cross-attention-based design against the alternative prefix-tuning
mechanism used in prior LM-based controllable music generation work
\cite{lin2023content_cocomulla,lan2024musicongen} in
Section~\ref{sec:exp_ablation} (w/ Prefix tuning).

At each modified layer $l$, the symbolic condition is first refined by a layer-specific
\emph{self-attention} block, denoted $\mathrm{SA}_l(\cdot)$, which allows the conditioning
to specialize to the level of abstraction expected by the corresponding decoder layer. The previous refined condition at layer $l-1$ is also fed into it to improve the expressivity:
\begin{equation}
    c_l \;=\; \mathrm{SA}_l(c + c_{l-1}) \;\in\; \mathbb{R}^{T \times D}.
\end{equation}
Note that $c_{l-1}$ is omitted at the first injecting layer $l=N-L+1$. Here and throughout, $\mathrm{SA}(\cdot)$ stands for a standard multi-head self-attention
block \cite{vaswani2017attention}.

The refined symbolic states $c_l$ are then merged into $h_l$ through a
\emph{cross-attention} block, which we denote $\mathrm{CA}_l(q, m)$: $q$ is the query
input and $m$ is the source input, from which the keys and the values are derived; the output has the same length
as $q$. Before $c_l$ is fed in as the source input, we apply rotary positional
embeddings (RoPE) \cite{su2024roformer} to it, so that the temporal positions of the symbolic
sequence are carried through into the keys and values together with their content. The
query input is the audio-side hidden states $h_l$.

A naive residual update of the form $\tilde{h}_l = h_l + \mathrm{CA}_l(h_l, c_l)$
admits the conditioning update at every position and feature dimension indiscriminately,
leaving no room for the model to decide where the symbolic information should actually be
incorporated; we observe in Section~\ref{sec:exp_ablation} (w/o Gate) that this
substantially degrades cover identifiability. Inspired by \cite{lin2023content_cocomulla},
we therefore introduce a \emph{learned gate} $g_l \in (-1, 1)^{S \times D}$ that lets the
model decide adaptively, for each position and feature dimension, how much symbolic
information to admit at layer $l$:
\begin{align}
    \tilde{h}_l & \;=\; h_l \;+\; g_l \;\odot\; \mathrm{CA}_l\bigl(h_l, c_l\bigr),
    \label{eq:inject}                                                                           \\
    g_l         & \;=\; \tanh\bigl(\mathrm{CA}^{\mathrm{gate}}_l\bigl(h_l, c_l\bigr)\bigr),
    \label{eq:gate}
\end{align}
where $\odot$ denotes elementwise multiplication, $\tanh$ bounds the gate to $(-1, 1)$, and
$\mathrm{CA}^{\mathrm{gate}}_l$ is a parameter-light cross-attention block that shares the
same $(q, m)$ convention as $\mathrm{CA}_l$ and likewise receives the RoPE-augmented
$c_l$ as its source input. We zero-initialize the output projection of
$\mathrm{CA}^{\mathrm{gate}}_l$, so that $g_l = \tanh(0) = 0$ at the start of training and
the augmented model exactly reproduces the pretrained backbone's behavior at
initialization.

\textbf{Partial conditioning via attention masking.}
The condition $c$ is built from the full-song lead sheet, yet \ours{} also supports
covers of arbitrary song segments through attention masking, naturally instantiating
the two canonical partial-generation modes. The user specifies a mask
interval via two scalars $\alpha, \beta \in [0, 1]$ with $\alpha \le \beta$, together
with a binary inversion flag $\rho \in \{0, 1\}$: $\rho{=}0$ keeps the interval as
the only conditioned region (\emph{outpainting}, in which the model is asked to
freely generate the rest of the song around a given segment), while $\rho{=}1$
keeps its complement (\emph{inpainting}, in which the model fills in the masked
interval given the surrounding conditioning context). These parameters induce a
set of \emph{active positions}
\begin{equation}
    \mathcal{A}(\alpha, \beta, \rho) =
    \begin{cases}
        \{\, t \in \{1, \dots, T\} : \alpha T < t \le \beta T \,\},                           & \rho = 0 \quad (\text{outpainting}), \\
        \{1, \dots, T\} \setminus \{\, t \in \{1, \dots, T\} : \alpha T < t \le \beta T \,\}, & \rho = 1 \quad (\text{inpainting}).
    \end{cases}
    \label{eq:active}
\end{equation}
Within $\mathrm{SA}_l$, self-attention is restricted to position pairs both lying in
$\mathcal{A}$; within $\mathrm{CA}_l$ and $\mathrm{CA}^{\mathrm{gate}}_l$, queries from
$h_l$ may only attend to keys and values whose source position lies in $\mathcal{A}$.
Positions outside $\mathcal{A}$ are thus ignored entirely by the injection pathway. The
defaults $(\alpha, \beta, \rho) = (0, 1, 0)$ recover the full-song setting; the three
masking modes are illustrated in Figure~\ref{fig:reprise}(d).

\subsection{Curriculum Training Strategy}
\label{sec:training}

We train \ours{} in three stages. Stages~1 and~2 establish the mapping from the lead
sheet to the audio at increasing scale; Stage~3 exposes the model to cross-version
pairings of content and style.

\textbf{Stage 1: Short-clip warmup.}
We first train on $30$-second segments. Short clips are cheap to optimize and let the
newly introduced modules pick up the basic mapping from the lead sheet to the audio
quickly, before the model is asked to handle longer-range structure. Skipping this
stage degrades both perceptual quality and cover identifiability, as we show in
Section~\ref{sec:exp_ablation} (w/o Warmup).

\textbf{Stage 2: Full-song training.}
We then continue training on full-length recordings, extending the mapping learned in
Stage~1 to song-scale temporal context. The existing controls $\ell, s$, the tonal-content
signal $c$, and the audio target are all drawn from the same song in Stage 1 and 2.

\textbf{Stage 3: Cross-cover finetuning.}
A cover song takes the tonal content of one rendition and re-renders it under the lyrics
and style of another. Stages~1 and~2 do not expose the model to this cross-version
pairing, since $c$ and $\ell, s$ are always drawn from the same recording. To close this
gap, we collect, for each song, a group of $K$ available renditions $\{v_1, \dots, v_K\}$,
and at each training step sample two indices $i, j \in \{1, \dots, K\}$:
the tonal-content signal $c$ is extracted from rendition $v_i$, while the remaining
controls $\ell, s$ and the audio target are taken from rendition $v_j$. The model is thus
explicitly trained to render the tonal content of $v_i$ under the lyrics and style of
$v_j$, which is precisely the cover-generation behavior we wish to evoke at inference.
Figure~\ref{fig:reprise}(c) contrasts this Stage~3 cross-version pairing with the
same-rendition setup of Stages~1 and~2. The contribution of this cross-cover pairing
to style controllability is quantified in Section~\ref{sec:exp_ablation}
(w/ SP).

\section{Experiments}
\label{sec:experiments}

This section describes the training setup of \ours{} and conducts evaluation on the full-song cover task and partial-reference cover task. Through objective and subjective
evaluation, we show that \ours{} surpasses representative open-source systems
and is on par with the strongest closed-source model. We conduct
ablation studies to show the necessity of each design choice.

\subsection{Training Configurations}
\label{sec:exp_training}

Following the data-construction protocol of recent work \cite{lisongecho,jiang2026muse},
we obtained approximately 200,000 synthetic songs from music generation providers. 
Each track is passed through an
automated annotation pipeline: vocals are isolated with Demucs \cite{defossez2019demucs}
and transcribed with Whisper \cite{radford2023robust_whisper}; section-level structure is
obtained by ensembling All-in-One \cite{kim2023all} and SongFormer \cite{hao2025songformer};
and free-form style tags produced by Qwen2-Audio \cite{chu2024qwen2} are formatted with
DeepSeek \cite{liu2024deepseek} into the tag vocabulary used by HeartMuLa
\cite{yang2026heartmula}. Lead sheet is transcribed using YourMT3 \cite{chang2024yourmt3+} for melody track and \cite{jiang2019large_chord} for chord progression. For Stage~3, we additionally curate a paired cross-cover set:
for each of 1{,}250 source songs we query music generation providers 
to synthesize 7 stylistically distinct covers, yielding 1{,}250 clusters of 8 interpretations
each (the source plus 7 covers). The full data construction pipeline for Stage~3 is described
in Appendix~\ref{app:stage3_pipeline}.

We build \ours{} on the 3B-parameter variant of HeartMuLa \cite{yang2026heartmula},
an LM-based music generation model with $N=28$ backbone transformer layers. Our Gated Adaptive Cross-Attention modules are inserted into the
last $L=8$ backbone layers. Stages~1 and~2 keep the backbone frozen and update only
the newly introduced parameters; Stage~3 additionally applies LoRA \cite{hu2022lora} to the last $L=8$ backbone layers and finetunes
on the cross-cover pairs. All experiments run on 16$\times$NVIDIA~A100 (80\,GB)
GPUs; per-stage hyperparameters are listed in Appendix~\ref{app:training_hparams}.

\subsection{Tasks and Baselines}
\label{sec:exp_tasks}
We evaluate \ours{} on two cover-generation tasks. Task~1 is full-song cover generation, where
the model is asked to generate a cover conditioned on the reference music, target style, and corresponding lyrics.
Task~2 is partial-reference cover generation. Under this setting, only a fragment of the reference music
is observed, along with the target style and lyrics for the whole song.
The model is asked to produce a full-song cover, preserving the tonal content of the observed part while
recreating the remainder. Instantiating the active-position mask $\mathcal{A}(\alpha, \beta, \rho)$ from
Eq.~\ref{eq:active}, Task~2 admits three setups---continuation, outpainting, and inpainting---detailed in Appendix~\ref{app:task_setup}.

We compare against five contemporary systems: SongEcho
\cite{lisongecho}, ACE-Step-1.5
\cite{gong2026ace}, YuE2 \cite{yuan2026yue2unifyingsymbolicaudio}, Mureka \cite{mureka_official} 
and Suno-v5.5 \cite{Suno_v5_5}. See Appendix~\ref{app:task_setup} for baseline configurations. Both Task~1 and Task~2 are evaluated objectively in
Section~\ref{sec:exp_objective}; for Task~2, objective metrics are averaged across the three setups, with per-setup metrics reported in Appendix~\ref{app:task_setup}. Task~1 additionally receives a
subjective evaluation in Section~\ref{sec:exp_subjective}. Suno-v5.5, Mureka and YuE2 do not directly support the partial-reference
setting and are therefore omitted from comparison in Task 2.

\subsection{Objective Evaluation}
\label{sec:exp_objective}

We prompt DeepSeek \cite{liu2024deepseek} to generate 32 sets of
lyrics and feed them to Suno-v5.5 \cite{Suno_v5_5} to obtain 32
reference songs, and prompt DeepSeek again to generate 8 distinct
style tags. Each model under evaluation is conditioned on a reference
song, its lyrics, and one style tag at a time, producing 8 stylistic
variants per reference song for a target total of 256 covers per model.
For the additional full-song baselines, YuE2 has all 256 outputs; Mureka
has 248 outputs because reference song 19 is absent under all eight styles.
Their results are averaged over available outputs within each style and then
across the eight styles.

We assess perceptual quality with two complementary pretrained
predictors. From AudioBox-Aesthetics \cite{tjandra2025meta_audiobox} we
report Production Quality (PQ), Production Complexity (PC), Content
Enjoyment (CE), and Content Usefulness (CU). From
SongEval \cite{yao2025songeval} we report Coherence (Coh.), Memorability
(Mem.), Naturalness (Nat.), Clarity (Cla.), and Musicality (Mus.).
For lyric controllability, we isolate the vocal track of each cover
with Demucs \cite{defossez2019demucs}, transcribe it with
Whisper \cite{radford2023robust_whisper}, and report Phoneme Error Rate (PER) against
the input lyrics. For style, we use the MuQ-MuLan music--text
contrastive model \cite{zhu2025muq}: MuQ-Sim is the cosine similarity
between the cover and its target style tag in the joint embedding
space, measuring fidelity to the requested style; MuQ-Dist is the
cosine distance between the cover and the source recording, measuring
stylistic deviation from the source. For cover identifiability, we
utilize cover song identification model CoverHunter \cite{coverhunter}: each cover serves as a query
against the database of 32 source songs, and we report top-1
retrieval accuracy (CSI-R@1) and the mean retrieval rank of the true
source (CSI-Rank).

\par\addvspace{8pt}
\begingroup
\captionsetup{type=table}
    \centering
    \begin{minipage}{\linewidth}
\setlength{\parskip}{0pt}
\caption{Objective evaluation on the full-song cover task (Task~1). $\uparrow$ /
        $\downarrow$ indicate higher / lower is better. \textbf{Bold} marks
        the best and \underline{underline} the second-best result in each
        column.}
    \label{tab:obj_main}
\fontsize{9}{11}\selectfont
\setlength{\tabcolsep}{3pt}
\renewcommand{\arraystretch}{1.08}
\noindent\textbf{AudioBox quality ($\uparrow$) \cite{tjandra2025meta_audiobox}}\par\smallskip
\begin{tabular*}{\textwidth}{@{\extracolsep{\fill}}lcccc@{}}
\toprule
Method & PQ & PC & CE & CU \\
\midrule
SongEcho \cite{lisongecho} & 7.88 & 6.11 & 7.26 & 7.54 \\
ACE-Step-1.5 \cite{gong2026ace} & 8.29 & 6.30 & 7.68 & 7.86 \\
Suno-v5.5 \cite{Suno_v5_5} & 8.30 & \textbf{6.44} & 7.66 & \underline{7.92} \\
Mureka~\cite{mureka_official} & 8.31 & \underline{6.39} & 7.61 & 7.82 \\
YuE2~\cite{yuan2026yue2unifyingsymbolicaudio} & \underline{8.33} & 6.30 & \textbf{7.71} & 7.91 \\
\ours{} & \textbf{8.34} & 6.34 & \underline{7.70} & \textbf{7.93} \\
\bottomrule
\end{tabular*}
\end{minipage}
\par\addvspace{8pt}
\begin{minipage}{\linewidth}
\setlength{\parskip}{0pt}
\captionsetup{type=table}
\caption*{Table~\ref{tab:obj_main} (continued).}
\fontsize{9}{11}\selectfont
\setlength{\tabcolsep}{3pt}
\renewcommand{\arraystretch}{1.08}
\noindent\textbf{SongEval aesthetics ($\uparrow$) \cite{yao2025songeval}}\par\smallskip
\begin{tabular*}{\textwidth}{@{\extracolsep{\fill}}lccccc@{}}
\toprule
Method & Coh. & Mem. & Nat. & Cla. & Mus. \\
\midrule
SongEcho \cite{lisongecho} & 3.70 & 3.57 & 3.37 & 3.47 & 3.46 \\
ACE-Step-1.5 \cite{gong2026ace} & 4.48 & 4.46 & 4.26 & 4.35 & 4.36 \\
Suno-v5.5 \cite{Suno_v5_5} & \textbf{4.61} & \textbf{4.60} & \textbf{4.42} & \textbf{4.52} & \textbf{4.51} \\
Mureka~\cite{mureka_official} & 4.28 & 4.24 & 4.05 & 4.12 & 4.11 \\
YuE2~\cite{yuan2026yue2unifyingsymbolicaudio} & 4.48 & 4.46 & 4.24 & 4.35 & 4.36 \\
\ours{} & \underline{4.59} & \underline{4.59} & \underline{4.34} & \underline{4.44} & \underline{4.47} \\
\bottomrule
\end{tabular*}
\end{minipage}
\par\addvspace{8pt}
\begin{minipage}{\linewidth}
\setlength{\parskip}{0pt}
\captionsetup{type=table}
\caption*{Table~\ref{tab:obj_main} (continued).}
\fontsize{9}{11}\selectfont
\setlength{\tabcolsep}{3pt}
\renewcommand{\arraystretch}{1.08}
\noindent\textbf{Lyrics, style, and cover identification}\par\smallskip
\begin{tabular*}{\textwidth}{@{\extracolsep{\fill}}lccccc@{}}
\toprule
Method & PER ($\downarrow$) & MuQ-Sim ($\uparrow$) & MuQ-Dist ($\uparrow$) & CSI-R@1 ($\uparrow$) & CSI-Rank ($\downarrow$) \\
\midrule
SongEcho \cite{lisongecho} & 0.26 & 0.38 & 0.39 & 0.88 & 1.55 \\
ACE-Step-1.5 \cite{gong2026ace} & 0.21 & 0.30 & 0.40 & 0.95 & 1.10 \\
Suno-v5.5 \cite{Suno_v5_5} & \textbf{0.11} & \textbf{0.50} & 0.33 & \underline{0.98} & 1.11 \\
Mureka~\cite{mureka_official} & 0.49 & \underline{0.48} & \underline{0.40} & \textbf{0.99} & \textbf{1.02} \\
YuE2~\cite{yuan2026yue2unifyingsymbolicaudio} & 0.22 & 0.45 & 0.32 & 0.96 & 1.06 \\
\ours{} & \underline{0.13} & 0.41 & \textbf{0.48} & 0.97 & \underline{1.05} \\
\bottomrule
\end{tabular*}
\end{minipage}
\par\endgroup\addvspace{8pt}

\par\addvspace{8pt}
\begingroup
\captionsetup{type=table}
    \centering
    \begin{minipage}{\linewidth}
\setlength{\parskip}{0pt}
\caption{Objective evaluation on the partial-reference cover task
        (Task~2), averaged over the three masking configurations
        (continuation, inpainting, outpainting). $\uparrow$ / $\downarrow$
        indicate higher / lower is better. \textbf{Bold} marks the best.}
    \label{tab:obj_partial}
\fontsize{9}{11}\selectfont
\setlength{\tabcolsep}{3pt}
\renewcommand{\arraystretch}{1.08}
\noindent\textbf{AudioBox quality ($\uparrow$) \cite{tjandra2025meta_audiobox}}\par\smallskip
\begin{tabular*}{\textwidth}{@{\extracolsep{\fill}}lcccc@{}}
\toprule
Method & PQ & PC & CE & CU \\
\midrule
SongEcho \cite{lisongecho} & 7.58 & 5.59 & 6.80 & 7.44 \\
ACE-Step-1.5 \cite{gong2026ace} & 8.26 & 5.91 & 7.66 & 7.84 \\
\ours{} & \textbf{8.33} & \textbf{6.31} & \textbf{7.69} & \textbf{7.92} \\
\bottomrule
\end{tabular*}
\end{minipage}
\par\addvspace{8pt}
\begin{minipage}{\linewidth}
\setlength{\parskip}{0pt}
\captionsetup{type=table}
\caption*{Table~\ref{tab:obj_partial} (continued).}
\fontsize{9}{11}\selectfont
\setlength{\tabcolsep}{3pt}
\renewcommand{\arraystretch}{1.08}
\noindent\textbf{SongEval aesthetics ($\uparrow$) \cite{yao2025songeval}}\par\smallskip
\begin{tabular*}{\textwidth}{@{\extracolsep{\fill}}lccccc@{}}
\toprule
Method & Coh. & Mem. & Nat. & Cla. & Mus. \\
\midrule
SongEcho \cite{lisongecho} & 3.14 & 3.04 & 2.85 & 2.87 & 2.92 \\
ACE-Step-1.5 \cite{gong2026ace} & 4.22 & 4.20 & 3.95 & 4.08 & 4.08 \\
\ours{} & \textbf{4.57} & \textbf{4.57} & \textbf{4.32} & \textbf{4.42} & \textbf{4.45} \\
\bottomrule
\end{tabular*}
\end{minipage}
\par\addvspace{8pt}
\begin{minipage}{\linewidth}
\setlength{\parskip}{0pt}
\captionsetup{type=table}
\caption*{Table~\ref{tab:obj_partial} (continued).}
\fontsize{9}{11}\selectfont
\setlength{\tabcolsep}{3pt}
\renewcommand{\arraystretch}{1.08}
\noindent\textbf{Lyrics, style, and cover identification}\par\smallskip
\begin{tabular*}{\textwidth}{@{\extracolsep{\fill}}lccccc@{}}
\toprule
Method & PER ($\downarrow$) & MuQ-Sim ($\uparrow$) & MuQ-Dist ($\uparrow$) & CSI-R@1 ($\uparrow$) & CSI-Rank ($\downarrow$) \\
\midrule
SongEcho \cite{lisongecho} & 0.55 & 0.25 & \textbf{0.63} & 0.41 & 5.94 \\
ACE-Step-1.5 \cite{gong2026ace} & 0.36 & 0.34 & 0.50 & 0.52 & 4.92 \\
\ours{} & \textbf{0.12} & \textbf{0.42} & 0.50 & \textbf{0.70} & \textbf{2.60} \\
\bottomrule
\end{tabular*}
\end{minipage}
\par\endgroup\addvspace{8pt}

Table~\ref{tab:obj_main} reports objective results on Task~1. Overall, \ours{} outperforms the open-source baselines on most metrics and performs on par with the closed-source commercial systems Suno-v5.5 and Mureka. For audio quality, \ours{} generates songs whose production quality and musicality are comparable to those of the commercial systems and generally better than those of the open-source baselines, suggesting that adding fine-grained controls does not compromise generation quality.
For controllability, \ours{} follows the given lyrics much more faithfully than the open-source baselines and nearly matches Suno-v5.5. In terms of style, it departs from the source arrangement more substantially than all baselines while maintaining reasonable fidelity to the target style, and the CSI results confirm that the generated songs remain recognizable as covers of the source song. In other words, \ours{} changes the style without losing the identity of the original song.

Table~\ref{tab:obj_partial} reports objective results on Task 2 averaged over the three masking configurations
(continuation, inpainting, outpainting); per-configuration breakdowns are
deferred to Appendix~\ref{app:partial_per_config}. When only a fragment of the reference music is exposed, the baselines suffer a sharp drop in cover identifiability, accompanied by varying degrees of degradation in lyric controllability and audio quality. \ours{}, in contrast, degrades far more gracefully and achieves the best performance on all metrics except MuQ-Dist, where SongEcho's higher value coincides with its near-total loss of identifiability and thus reflects drift from the source rather than controlled stylization. This further demonstrates the flexibility of our control scheme.

\subsection{Subjective Evaluation}
\label{sec:exp_subjective}

\paragraph{Listening study.}
We complement the objective evaluation with a blind listening study. From the
test set introduced in Section~\ref{sec:exp_objective} we randomly select
8~source songs, each paired with a target style tag and the corresponding
lyrics. For each song we generate a cover with all four systems
(\ours{}, ACE-Step-1.5, SongEcho, and Suno-v5.5), shuffle their order, and
ask each rater to score every cover on four 5-point Likert
scales: \emph{Identity Preserveness}, \emph{Style Controllability},
\emph{Creativeness}, and \emph{Overall Musicality}.
Each participant was assigned a random subset of 5 of the 8 songs.
We received 34 saved submissions. After excluding submissions with no complete
case and dropping incomplete case blocks, the analysis retained 22 participants
self-reporting their musical background as
amateur (11), intermediate (6), or professional (5), yielding 95 complete
participant--case blocks and 1,520 ratings in total. Partial completion resulted
in 4.32 complete cases per retained participant on average.

Table~\ref{tab:subj_main} reports the per-metric mean and 95\% confidence
interval across all (participant, song) pairs. Under paired block-level
Wilcoxon tests, \ours{} significantly outperforming open-sourced baselines on Style Controllability, Creativeness, and Overall Musicality and is on par with Suno-v5.5.
On Identity Preserveness, \ours{} significantly outperforms SongEcho and shows
no statistically significant difference from ACE-Step-1.5, while Suno-v5.5 remains
significantly higher. The full questionnaire instrument, scoring rubrics, and detailed instructions
are detailed in Appendix~\ref{app:subjective_protocol}.

\begin{table}[t]
    \centering
    \caption{Listening study: mean $\pm$ 95\% confidence interval on each
        metric (5-point Likert; higher is better). \textbf{Bold} marks the best
        and \underline{underline} the second-best result in each column.}
    \label{tab:subj_main}
    \resizebox{\textwidth}{!}{
        \begin{tabular}{lcccc}
            \toprule
            Method                          & Identity Preserveness ($\uparrow$) & Style Controllability ($\uparrow$) & Creativeness ($\uparrow$)   & Overall Musicality ($\uparrow$) \\
            \midrule
            SongEcho \cite{lisongecho}      & $2.84 \pm 0.19$                    & $2.84 \pm 0.23$                    & $2.56 \pm 0.21$             & $2.28 \pm 0.21$                 \\
            ACE-Step-1.5 \cite{gong2026ace} & \underline{$3.53 \pm 0.21$}        & $2.93 \pm 0.23$                    & $3.03 \pm 0.21$             & $3.31 \pm 0.22$                 \\
            Suno-v5.5 \cite{Suno_v5_5}      & $\mathbf{3.81 \pm 0.19}$           & $\mathbf{3.72 \pm 0.21}$           & \underline{$3.51 \pm 0.20$} & $\mathbf{3.88 \pm 0.21}$        \\
            \midrule
            \ours{}                         & $3.45 \pm 0.20$                    & \underline{$3.71 \pm 0.25$}        & $\mathbf{3.65 \pm 0.22}$    & \underline{$3.75 \pm 0.22$}     \\
            \bottomrule
        \end{tabular}
    }
\end{table}

\paragraph{Expert interview.}
The Likert questionnaire reveals \emph{how good} each system is but not
\emph{why}; to probe the latter, we run a structured expert-interview protocol
and report an R\&B case study. We first sample one genre uniformly at random
from a catalogue of common popular-music styles and, without playing any audio,
ask the expert to enumerate from prior experience the salient attributes a
faithful instance of that genre should exhibit, each with a short
justification. We then sample one source song from the test set, present the
four blinded covers, and ask the expert to rate and comment on each version
against the elicited attributes. Finally, the commentary is fed to a language
model to produce a concise summary verdict. The full protocol, the
sampled genre, the elicited attributes, the organized expert record, and the
summarization prompt are reproduced in Appendix~\ref{app:expert_interviews},
where Versions~A, B, C, D in the blinded record correspond to ACE-Step-1.5,
SongEcho, \ours{}, and Suno-v5.5, respectively. The summary verdict reads as
follows:

\begin{quote}
    \itshape
    Under the expert's two pre-specified R\&B criteria—rhythm/groove and vocal melodic development—the four blinded versions rank D > C > B > A, with A deviating most toward electronic pop, B and C progressively closer to R\&B's cyclic vocal phrasing, and D judged most canonical thanks to its natural groove and pop-R\&B vocal timbre, melismas, and phrase endings.
\end{quote}

\subsection{Ablation Studies}
\label{sec:exp_ablation}
\par\addvspace{8pt}
\begingroup
\captionsetup{type=table}
    \centering
    \begin{minipage}{\linewidth}
\setlength{\parskip}{0pt}
\caption{Ablation studies on the full-song cover task. The first row
        is the full \ours{}; rows below ablate one component at a time,
        grouped by \emph{training schedule}, \emph{condition signal}, and
        \emph{architecture}. $\uparrow$ / $\downarrow$ indicate higher /
        lower is better.}
    \label{tab:abl}
\fontsize{8.5}{10}\selectfont
\setlength{\tabcolsep}{3pt}
\renewcommand{\arraystretch}{1.08}
\noindent\textbf{AudioBox quality ($\uparrow$) \cite{tjandra2025meta_audiobox}}\par\smallskip
\begin{tabular*}{\textwidth}{@{\extracolsep{\fill}}lcccc@{}}
\toprule
Variant & PQ & PC & CE & CU \\
\midrule
\ours{}-small & 8.28 & 6.36 & 7.67 & 7.89 \\
\multicolumn{5}{l}{\textit{Training schedule}} \\
\quad w/o Warmup & 8.07 & 6.13 & 7.44 & 7.72 \\
\quad w/ SP & 8.24 & 6.43 & 7.63 & 7.87 \\
\multicolumn{5}{l}{\textit{Condition signal}} \\
\quad w/o BPM-norm & 8.21 & 6.36 & 7.61 & 7.84 \\
\quad w/o BPM-norm (SP.) & 8.19 & 6.40 & 7.59 & 7.82 \\
\quad w/ $F_0$ & 8.10 & 6.46 & 7.55 & 7.74 \\
\quad w/ $F_0$ (SP.) & 8.12 & 6.45 & 7.56 & 7.76 \\
\quad w/o Chord & 7.97 & 6.30 & 7.39 & 7.65 \\
\multicolumn{5}{l}{\textit{Architecture}} \\
\quad w/o Gate & 8.26 & 6.38 & 7.62 & 7.88 \\
\quad w/ Prefix tuning & 7.97 & 5.81 & 7.37 & 7.67 \\
\bottomrule
\end{tabular*}
\end{minipage}
\par\addvspace{8pt}
\begin{minipage}{\linewidth}
\setlength{\parskip}{0pt}
\captionsetup{type=table}
\caption*{Table~\ref{tab:abl} (continued).}
\fontsize{8.5}{10}\selectfont
\setlength{\tabcolsep}{3pt}
\renewcommand{\arraystretch}{1.08}
\noindent\textbf{SongEval aesthetics ($\uparrow$) \cite{yao2025songeval}}\par\smallskip
\begin{tabular*}{\textwidth}{@{\extracolsep{\fill}}lccccc@{}}
\toprule
Variant & Coh. & Mem. & Nat. & Cla. & Mus. \\
\midrule
\ours{}-small & 4.58 & 4.56 & 4.33 & 4.42 & 4.45 \\
\multicolumn{6}{l}{\textit{Training schedule}} \\
\quad w/o Warmup & 4.41 & 4.40 & 4.13 & 4.24 & 4.23 \\
\quad w/ SP & 4.57 & 4.55 & 4.33 & 4.41 & 4.44 \\
\multicolumn{6}{l}{\textit{Condition signal}} \\
\quad w/o BPM-norm & 4.56 & 4.55 & 4.33 & 4.40 & 4.43 \\
\quad w/o BPM-norm (SP.) & 4.48 & 4.45 & 4.23 & 4.30 & 4.35 \\
\quad w/ $F_0$ & 4.45 & 4.41 & 4.19 & 4.24 & 4.41 \\
\quad w/ $F_0$ (SP.) & 4.44 & 4.38 & 4.16 & 4.23 & 4.39 \\
\quad w/o Chord & 4.35 & 4.32 & 4.07 & 4.16 & 4.16 \\
\multicolumn{6}{l}{\textit{Architecture}} \\
\quad w/o Gate & 4.59 & 4.58 & 4.34 & 4.44 & 4.46 \\
\quad w/ Prefix tuning & 4.34 & 4.35 & 4.08 & 4.16 & 4.18 \\
\bottomrule
\end{tabular*}
\end{minipage}
\par\addvspace{8pt}
\begin{minipage}{\linewidth}
\setlength{\parskip}{0pt}
\captionsetup{type=table}
\caption*{Table~\ref{tab:abl} (continued).}
\fontsize{8.5}{10}\selectfont
\setlength{\tabcolsep}{3pt}
\renewcommand{\arraystretch}{1.08}
\noindent\textbf{Lyrics, style, and cover identification}\par\smallskip
\begin{tabular*}{\textwidth}{@{\extracolsep{\fill}}lccccc@{}}
\toprule
Variant & PER ($\downarrow$) & MuQ-Sim ($\uparrow$) & MuQ-Dist ($\uparrow$) & CSI-R@1 ($\uparrow$) & CSI-Rank ($\downarrow$) \\
\midrule
\ours{}-small & 0.14 & 0.38 & 0.49 & 0.98 & 1.04 \\
\multicolumn{6}{l}{\textit{Training schedule}} \\
\quad w/o Warmup & 0.16 & 0.40 & 0.54 & 0.11 & 9.75 \\
\quad w/ SP & 0.16 & 0.35 & 0.45 & 0.98 & 1.04 \\
\multicolumn{6}{l}{\textit{Condition signal}} \\
\quad w/o BPM-norm & 0.14 & 0.36 & 0.51 & 0.73 & 2.70 \\
\quad w/o BPM-norm (SP.) & 0.18 & 0.34 & 0.47 & 0.99 & 1.04 \\
\quad w/ $F_0$ & 0.20 & 0.39 & 0.52 & 0.33 & 5.20 \\
\quad w/ $F_0$ (SP.) & 0.20 & 0.38 & 0.51 & 0.40 & 5.55 \\
\quad w/o Chord & 0.20 & 0.38 & 0.50 & 0.46 & 4.86 \\
\multicolumn{6}{l}{\textit{Architecture}} \\
\quad w/o Gate & 0.14 & 0.38 & 0.52 & 0.86 & 1.74 \\
\quad w/ Prefix tuning & 0.43 & 0.36 & 0.58 & 0.09 & 11.58 \\
\bottomrule
\end{tabular*}
\end{minipage}
\par\endgroup\addvspace{8pt}

To isolate the contribution of each design choice in \ours{}, we
ablate three axes of the system in turn: (i) the \emph{training
    schedule}, (ii) the \emph{condition signal}, and (iii) the
\emph{architecture} used to inject that signal into the backbone.
The metrics used in ablation studies share the same definition as described in Section~\ref{sec:exp_objective}
and are reported in Table~\ref{tab:abl}. Due to limited compute, we
retrain on a 70k-song subset (denoted \ours{}-small) and use the
same data scale for every variant.

\paragraph{Training schedule.}
We first ablate the short-clip warmup (Stage~1). The variant
w/o Warmup skips Stage~1 and starts training directly from
Stage~2, with one extra epoch in Stage~2 to match the total number
of training steps. Without the warmup, the newly introduced modules
fail to align with the backbone and $c$ is not effectively
injected. The output is no longer identifiable
as a cover, and the broken injection collaterally drags down
perceptual quality. We next ablate the cross-cover pairing in Stage~3, where $c$ is
drawn from a different rendition than the one supplying
$\ell, s, a$. The variant w/ SP (self-pair) reverts Stage~3
to the same-rendition setup of Stages~1 and~2, with extra epochs
compensating for the reduced number of $(c, \ell, s, a)$ tuples.
Style controllability deteriorates while everything else is
preserved, confirming that the cross-cover stage is what unlocks
the model's stylistic flexibility.

\paragraph{Condition signal.}
Our symbolic tonal-content signal is BPM-normalized to a canonical
$\mathrm{BPM}{=}120$, with the intent of relaxing the otherwise
strict frame-wise correspondence between $c$ and the audio so as
to enhance style controllability without compromising content
control. To verify this, we replace $c$ with a non-BPM-normalized
symbolic sequence whose length and frame rate match the audio
token stream, and train two variants: w/o BPM-norm (the full
three-stage schedule including the cross-cover Stage~3) and
w/o BPM-norm (SP.) (Stage~3 restricted to self-pairs). The two
together expose a content/style tradeoff that is structural rather
than incidental. Once $c$ is strictly frame-wise aligned with the
audio, the self-pair regime is fully consistent with the alignment
and yields strong cover identifiability at the cost of style
flexibility, whereas the cross-cover regime pairs the condition
of one rendition with the audio of another, and a frame-wise
signal becomes too brittle to small tempo or duration mismatches,
so cover identifiability collapses instead. BPM normalization
loosens the temporal alignment just enough to keep the content
needed for cover identifiability while leaving the slack required
by style control.

\Needspace{3\baselineskip}
We note that SongEcho \cite{lisongecho} uses a frame-wise
$F_0$ sequence as its content control. Substituting our symbolic
signal with the same $F_0$ representation gives the analogous pair
of variants w/ $F_0$ and w/ $F_0$ (SP.), and the same content/style
tradeoff between the two reappears. Beyond the tradeoff, $F_0$
based injection is uniformly weaker than the symbolic counterpart,
which we attribute to two intrinsic properties of $F_0$: being
continuous rather than discretized to per-pitch values, it is
harder to internalize after embedding and thus less amenable to
injection; and carrying only the vocal melody, it is a sparser
and less informative signal in the first place.

The variant w/o Chord drops the chord progression and keeps only
the melody, and ineffective injection
emerges. In popular music the chord progression carries the
harmonic \emph{skeleton} of the song and is far more informative
than the vocal melody alone, and we believe removing it leaves $c$ too sparse for
the model to reliably exploit.

Taken together, these results justify our choice of a
BPM-normalized symbolic lead sheet (melody plus chord progression)
as the tonal-content signal: among the design alternatives we
considered, it best balances content control (cover
identifiability) and style control.

\paragraph{Architecture.}
We first remove the learned gate $g_l$ in Equation~\ref{eq:inject},
reducing the injection to a plain residual cross-attention
$\tilde{h}_l = h_l + \mathrm{CA}_l(h_l, c_l)$; we denote this
variant w/o Gate. Cover identifiability suffers, consistent with
our hypothesis that the adaptive gate is precisely the mechanism
through which the model selects which positions and feature
dimensions of the conditioning update should be admitted into
$h_l$. Without it, the symbolic information is broadcast
indiscriminately and the model loses the freedom to bind the
condition where it is actually useful. We next replace cross-attention-based injection with a
prefix-tuning mechanism, previously shown to be effective for
adapting LM-based music generators to new
controls~\cite{lin2023content_cocomulla,lan2024musicongen}; we
denote this variant w/ Prefix tuning. Here condition injection
essentially fails, and prefix tuning even disrupts the
lyrics-to-music mapping already established by the foundation
model, as reflected in the PER blow-up. We attribute this to the
fact that our backbone already consumes lyrics and style as prefix
prompts, and forcing
an additional newly introduced control interferes with the autoregressive mapping the model previously
learned.

Overall, GACA is a sound and novel choice for condition injection
on top of an LM-based music foundation model, compared with the
alternatives we examined.

\section{Discussion and Limitations}
\label{sec:discussion}

\ours{} formulates cover song generation as lead-sheet-controlled music generation. However, we acknowledge that the notion of a cover song is inherently open-ended: what should remain invariant across versions can be defined at different granularities, ranging from strictly preserving the lead sheet, to allowing re-harmonization in a jazz arrangement, to borrowing only a few motifs from the original as inspiration. Our current formulation fixes this boundary at the lead-sheet level. To better support human-AI co-creation in real-world scenarios, future work could offer more intuitive controls that let users adjust how closely the generated song adheres to the original.

Copyright protection for AI-generated content remains an open concern, and it is especially relevant to cover song generation, which by nature takes existing songs as input. We respect the intellectual property of human artists. All training data used in this work are synthetic and were obtained from music generation providers in accordance with their terms of service. Since generated covers may constitute derivative works of the original songs, users are responsible for obtaining the necessary permissions from rights holders before distributing or commercially using such outputs.

\section{Conclusion}
\label{sec:conclusion}

We present \ours{}, a lead-sheet-conditioned song generation system for
flexible cover generation that conditions a
pretrained text-to-song foundation model on a BPM-normalized symbolic
lead sheet via a gated cross-attention module. \ours{} is
alignment-free, supporting flexible tempo and
partial-reference generation in a single unified model. Objective
metrics, listening-study questionnaires, and a structured
expert-interview protocol jointly show that \ours{} outperforms
open-source cover baselines and is broadly on par with the leading
closed-source commercial system Suno-v5.5.

\FloatBarrier
\Needspace{10\baselineskip}
\appendix
\section*{Appendix}
\FloatBarrier
\section{Per-Stage Training Hyperparameters}
\label{app:training_hparams}

Table~\ref{tab:training_hparams} reports the per-stage training hyperparameters.

\begin{table}[!htbp]
    \centering
    \caption{Per-stage training hyperparameters. GACA: gated adaptive cross-attention. Entries marked $\star$ follow the dynamic mask schedule described below.}
    \label{tab:training_hparams}
    \setlength{\tabcolsep}{5pt}
    \renewcommand{\arraystretch}{1.15}
    \begin{tabular*}{\textwidth}{@{\extracolsep{\fill}}l cccc}
        \toprule
        Hyperparameter                      & Stage~1            & Stage~2 (w/o mask) & Stage~2 (w/ mask)  & Stage~3            \\
        \midrule
        Learning rate                       & $2{\times}10^{-4}$ & $2{\times}10^{-4}$ & $2{\times}10^{-4}$ & $2{\times}10^{-5}$ \\
        Epochs                              & 3                  & 2                  & 2                  & 2                  \\
        Avg.\ bs/GPU            & 9                  & 1                  & 1                  & 1                  \\
        $(\alpha, \beta, \rho)$ & $(0.0, 1.0, 0.0)$  & $(0.0, 1.0, 0.0)$  & $\star$            & $\star$            \\
        Trainable modules                   & GACA        & GACA        & GACA        & GACA \& LoRA                \\
        Trainable params                    & ${\sim}300$M       & ${\sim}300$M       & ${\sim}300$M       & ${\sim}400$M         \\
        Time per epoch                      & $0.5$\,h           & ${\sim}5$\,h       & ${\sim}5$\,h       & ${\sim}2$\,h       \\
        \midrule
        Optimizer                           & \multicolumn{4}{c}{AdamW}                                                       \\
        LR scheduler                        & \multicolumn{4}{c}{warmup followed by cosine annealing}                         \\
        Gradient accum.               & \multicolumn{4}{c}{4}                                                           \\
        GPUs                                & \multicolumn{4}{c}{16}                                                          \\
        \bottomrule
    \end{tabular*}
\end{table}

\paragraph{Mask configuration $(\alpha, \beta, \rho)$.}
$\alpha$, $\beta$, and $\rho$ denote the start ratio, end ratio, and invert ratio of the masked window. For Stage~1 and Stage~2 (w/o mask), the configuration $(0.0, 1.0, 0.0)$ corresponds to the default full-song setting. For entries marked $\star$ (Stage~2 w/ mask and Stage~3), we fix $\rho = 0.5$ and sample $\alpha, \beta$ per call: a window-length ratio $r$ is either fixed or drawn from a curriculum range $r \sim \mathcal{U}(r_{\min}, r_{\max})$; then $\alpha \sim \mathcal{U}(0, 1 - r)$ and $\beta = \alpha + r$.

\FloatBarrier
\section{Task Setup and Baseline Configurations}
\label{app:task_setup}

The partial-reference cover task (Task~2; Section~\ref{sec:exp_tasks})
exposes only a fragment of the reference recording to the model. For each of the
$32$ reference pieces $i=1,2,\dots,32$, we manually label the start and end
$\alpha_i, \beta_i$ of the first chorus. The three setups---continuation, outpainting,
and inpainting---are then specified by instantiating the active-position mask
$\mathcal{A}(\alpha, \beta, \rho)$ from Eq.~\ref{eq:active} as follows.
\emph{(i)}~\textbf{Continuation} [$\mathcal{A}(0, \alpha_i, 0)$]: the prefix from the
song's start through the section preceding the first chorus is observed and the
remainder is generated.
\emph{(ii)}~\textbf{Outpainting} [$\mathcal{A}(\alpha_i, \beta_i, 0)$]: only the first
chorus is observed and the rest of the song (preceding and following) is generated
outward from it.
\emph{(iii)}~\textbf{Inpainting} [$\mathcal{A}(\alpha_i, \beta_i, 1)$]: everything
except the first chorus is observed, and the model is asked to fill in the missing chorus.

For each configuration, the observed fragment supplies the tonal-content
condition for the corresponding region, while the unobserved region is
generated freely under the target lyrics and style tag.

For SongEcho \cite{lisongecho}, we refer to its official implementation\footnote{\url{https://github.com/lsfhuihuiff/SongEcho_ICLR2026}}.
For inference on Task~1, we extract the f0 sequence using its provided RMVPE model and follow its official
inference Python script. We use its released checkpoint with all other parameters set to default. For Task~2,
the SongEcho paper claims support for inpainting and outpainting, also showcased on their demo page\footnote{\url{https://vvanonymousvv.github.io/SongEcho_updated/}};
however, the relevant instructions and inference script for this setting are not found in their open-sourced code. We therefore
refer to its description in the original paper, which we quote here: ``Our method supports
inpainting and outpainting via a simple masking strategy'' (Sec.~5.3.2 of the SongEcho paper), and implement a masking strategy based
on our understanding: for a given reference music, the observed part is kept as the condition, and the rest is masked out as zero silence.

For ACE-Step~1.5 \cite{gong2026ace}, we refer to its official implementation\footnote{\url{https://github.com/ace-step/ACE-Step-1.5}}. We use its turbo version.
We refer to its official documentation for cover song generation\footnote{\url{https://github.com/ace-step/ACE-Step-1.5/blob/main/docs/en/INFERENCE.md\#2-cover}}.
Besides lyrics and style prompts, the influence of the reference music is controlled by the parameter \texttt{audio\_cover\_strength}.
When this parameter is set too high ($\geq 0.4$), we empirically find that the model copies the original song; when set too low ($\leq 0.2$),
it cannot preserve the tonal content and therefore cannot be identified as a cover given the reference.
To ensure a fair comparison and a genuine baseline performance, we set \texttt{audio\_cover\_strength} to $0.3$. We additionally report ACE-Step~1.5's
objective metrics in Table \ref{tab:ace} while iterating through \texttt{audio\_cover\_strength} $\lambda$ from $0.1$ to $0.8$.
All the metrics converge as $\lambda \ge 0.4$, with 100 percent accuracy for cover song identification model. It is exactly caused by the model copying the original song and producing identical inference results.
While as $\lambda = 0.1$ or $0.2$, the accuracy of cover song identification is low, indicating that the model can not preserve the necessary tonal content that a cover version should keep invariant.
For Task~2, ACE-Step~1.5 does not natively support our setting; however, this can be achieved by cascading its cover and repaint features.
Cascading approach: for a reference music, split it into two parts---the partial-condition part and the remaining part. For the partial-condition part,
we separately use the cover feature to generate a cover, and we substitute the original portion with the generated part. We then feed the whole
song (partial-condition part replaced by the cover result while the remaining stays the same) into the model and ask it to repaint the rest.

\paragraph{Additional full-song baselines.}
Mureka and YuE2 are included only in the Task~1 objective comparison.
They use the reference-song, lyric, and target-style evaluation setup described
in Section~\ref{sec:exp_objective}. YuE2 contributes 256 outputs; Mureka
contributes 248 because source song 19 is absent under all eight target styles.
Scores are averaged over available outputs within each style, then across
the eight styles. Thus, Mureka's scores use a smaller reference set and should
be interpreted with this coverage difference in mind. Neither system is
included in Task~2, the four-system listening study, or the expert case study.

\par\addvspace{8pt}
\begingroup
\captionsetup{type=table}
    \centering
    \begin{minipage}{\linewidth}
\setlength{\parskip}{0pt}
\caption{Objective evaluation results under different audio cover control strengths $\lambda$ for ACE-Step 1.5.}
    \label{tab:ace}
\fontsize{9}{11}\selectfont
\setlength{\tabcolsep}{3pt}
\renewcommand{\arraystretch}{1.08}
\noindent\textbf{AudioBox quality ($\uparrow$) \cite{tjandra2025meta_audiobox}}\par\smallskip
\begin{tabular*}{\textwidth}{@{\extracolsep{\fill}}lcccc@{}}
\toprule
Method & PQ & PC & CE & CU \\
\midrule
ACE-Step-1.5 $\lambda=0.1$ & 8.25 & 5.58 & 7.64 & 7.80 \\
ACE-Step-1.5 $\lambda=0.2$ & 8.23 & 5.87 & 7.64 & 7.78 \\
ACE-Step-1.5 $\lambda=0.3$ & 8.28 & 6.30 & 7.67 & 7.86 \\
ACE-Step-1.5 $\lambda=0.4$ & 8.28 & 6.40 & 7.64 & 7.87 \\
ACE-Step-1.5 $\lambda=0.5$ & 8.27 & 6.43 & 7.61 & 7.86 \\
ACE-Step-1.5 $\lambda=0.6$ & 8.27 & 6.42 & 7.61 & 7.85 \\
ACE-Step-1.5 $\lambda=0.7$ & 8.25 & 6.43 & 7.60 & 7.84 \\
ACE-Step-1.5 $\lambda=0.8$ & 8.25 & 6.44 & 7.59 & 7.83 \\
\bottomrule
\end{tabular*}
\end{minipage}
\par\addvspace{8pt}
\begin{minipage}{\linewidth}
\setlength{\parskip}{0pt}
\captionsetup{type=table}
\caption*{Table~\ref{tab:ace} (continued).}
\fontsize{9}{11}\selectfont
\setlength{\tabcolsep}{3pt}
\renewcommand{\arraystretch}{1.08}
\noindent\textbf{SongEval aesthetics ($\uparrow$) \cite{yao2025songeval}}\par\smallskip
\begin{tabular*}{\textwidth}{@{\extracolsep{\fill}}lccccc@{}}
\toprule
Method & Coh. & Mem. & Nat. & Cla. & Mus. \\
\midrule
ACE-Step-1.5 $\lambda=0.1$ & 4.21 & 4.17 & 3.93 & 4.08 & 4.07 \\
ACE-Step-1.5 $\lambda=0.2$ & 4.28 & 4.25 & 4.03 & 4.14 & 4.17 \\
ACE-Step-1.5 $\lambda=0.3$ & 4.48 & 4.46 & 4.25 & 4.34 & 4.36 \\
ACE-Step-1.5 $\lambda=0.4$ & 4.51 & 4.49 & 4.29 & 4.37 & 4.38 \\
ACE-Step-1.5 $\lambda=0.5$ & 4.51 & 4.50 & 4.29 & 4.38 & 4.38 \\
ACE-Step-1.5 $\lambda=0.6$ & 4.51 & 4.49 & 4.28 & 4.38 & 4.38 \\
ACE-Step-1.5 $\lambda=0.7$ & 4.51 & 4.49 & 4.28 & 4.38 & 4.38 \\
ACE-Step-1.5 $\lambda=0.8$ & 4.51 & 4.50 & 4.27 & 4.38 & 4.37 \\
\bottomrule
\end{tabular*}
\end{minipage}
\par\addvspace{8pt}
\begin{minipage}{\linewidth}
\setlength{\parskip}{0pt}
\captionsetup{type=table}
\caption*{Table~\ref{tab:ace} (continued).}
\fontsize{9}{11}\selectfont
\setlength{\tabcolsep}{3pt}
\renewcommand{\arraystretch}{1.08}
\noindent\textbf{Lyrics, style, and cover identification}\par\smallskip
\begin{tabular*}{\textwidth}{@{\extracolsep{\fill}}lccccc@{}}
\toprule
Method & PER ($\downarrow$) & MuQ-Sim ($\uparrow$) & MuQ-Dist ($\uparrow$) & CSI-R@1 ($\uparrow$) & CSI-Rank ($\downarrow$) \\
\midrule
ACE-Step-1.5 $\lambda=0.1$ & 0.37 & 0.39 & 0.56 & 0.08 & 12.40 \\
ACE-Step-1.5 $\lambda=0.2$ & 0.29 & 0.37 & 0.50 & 0.25 & 8.03 \\
ACE-Step-1.5 $\lambda=0.3$ & 0.21 & 0.30 & 0.40 & 0.95 & 1.10 \\
ACE-Step-1.5 $\lambda=0.4$ & 0.13 & 0.27 & 0.36 & 1.00 & 1.00 \\
ACE-Step-1.5 $\lambda=0.5$ & 0.11 & 0.27 & 0.34 & 1.00 & 1.00 \\
ACE-Step-1.5 $\lambda=0.6$ & 0.12 & 0.27 & 0.34 & 1.00 & 1.00 \\
ACE-Step-1.5 $\lambda=0.7$ & 0.13 & 0.27 & 0.33 & 1.00 & 1.00 \\
ACE-Step-1.5 $\lambda=0.8$ & 0.12 & 0.27 & 0.33 & 1.00 & 1.00 \\
\bottomrule
\end{tabular*}
\end{minipage}
\par\endgroup\addvspace{8pt}

\FloatBarrier
\section{Partial-Reference Cover: Per-Configuration Results}
\label{app:partial_per_config}

Table~\ref{tab:obj_partial} in the main text reports objective scores
averaged over the three masking configurations. Tables~\ref{tab:partial_continuation}--\ref{tab:partial_inpainting}
report the per-configuration breakdown. Among each configuration, \ours{} achieves superior results over baselines.

\par\addvspace{8pt}
\begingroup
\captionsetup{type=table}
    \centering
    \begin{minipage}{\linewidth}
\setlength{\parskip}{0pt}
\caption{Partial-reference cover, \emph{continuation} configuration.
        Same metrics as Table~\ref{tab:obj_main}.}
    \label{tab:partial_continuation}
\fontsize{9}{11}\selectfont
\setlength{\tabcolsep}{3pt}
\renewcommand{\arraystretch}{1.08}
\noindent\textbf{AudioBox quality ($\uparrow$) \cite{tjandra2025meta_audiobox}}\par\smallskip
\begin{tabular*}{\textwidth}{@{\extracolsep{\fill}}lcccc@{}}
\toprule
Method & PQ & PC & CE & CU \\
\midrule
SongEcho \cite{lisongecho} & 7.49 & 5.44 & 6.63 & 7.43 \\
ACE-Step-1.5 \cite{gong2026ace} & 8.26 & 5.86 & 7.65 & 7.83 \\
\ours{} & 8.33 & 6.33 & 7.71 & 7.93 \\
\bottomrule
\end{tabular*}
\end{minipage}
\par\addvspace{8pt}
\begin{minipage}{\linewidth}
\setlength{\parskip}{0pt}
\captionsetup{type=table}
\caption*{Table~\ref{tab:partial_continuation} (continued).}
\fontsize{9}{11}\selectfont
\setlength{\tabcolsep}{3pt}
\renewcommand{\arraystretch}{1.08}
\noindent\textbf{SongEval aesthetics ($\uparrow$) \cite{yao2025songeval}}\par\smallskip
\begin{tabular*}{\textwidth}{@{\extracolsep{\fill}}lccccc@{}}
\toprule
Method & Coh. & Mem. & Nat. & Cla. & Mus. \\
\midrule
SongEcho \cite{lisongecho} & 3.06 & 2.98 & 2.76 & 2.77 & 2.83 \\
ACE-Step-1.5 \cite{gong2026ace} & 4.19 & 4.17 & 3.91 & 4.05 & 4.04 \\
\ours{} & 4.59 & 4.59 & 4.33 & 4.44 & 4.47 \\
\bottomrule
\end{tabular*}
\end{minipage}
\par\addvspace{8pt}
\begin{minipage}{\linewidth}
\setlength{\parskip}{0pt}
\captionsetup{type=table}
\caption*{Table~\ref{tab:partial_continuation} (continued).}
\fontsize{9}{11}\selectfont
\setlength{\tabcolsep}{3pt}
\renewcommand{\arraystretch}{1.08}
\noindent\textbf{Lyrics, style, and cover identification}\par\smallskip
\begin{tabular*}{\textwidth}{@{\extracolsep{\fill}}lccccc@{}}
\toprule
Method & PER ($\downarrow$) & MuQ-Sim ($\uparrow$) & MuQ-Dist ($\uparrow$) & CSI-R@1 ($\uparrow$) & CSI-Rank ($\downarrow$) \\
\midrule
SongEcho \cite{lisongecho} & 0.64 & 0.19 & 0.70 & 0.25 & 7.59 \\
ACE-Step-1.5 \cite{gong2026ace} & 0.36 & 0.35 & 0.51 & 0.38 & 6.18 \\
\ours{} & 0.11 & 0.42 & 0.51 & 0.63 & 3.15 \\
\bottomrule
\end{tabular*}
\end{minipage}
\par\endgroup\addvspace{8pt}

\par\addvspace{8pt}
\begingroup
\captionsetup{type=table}
    \centering
    \begin{minipage}{\linewidth}
\setlength{\parskip}{0pt}
\caption{Partial-reference cover, \emph{outpainting} configuration.
        Same metrics as Table~\ref{tab:obj_main}.}
    \label{tab:partial_outpainting}
\fontsize{9}{11}\selectfont
\setlength{\tabcolsep}{3pt}
\renewcommand{\arraystretch}{1.08}
\noindent\textbf{AudioBox quality ($\uparrow$) \cite{tjandra2025meta_audiobox}}\par\smallskip
\begin{tabular*}{\textwidth}{@{\extracolsep{\fill}}lcccc@{}}
\toprule
Method & PQ & PC & CE & CU \\
\midrule
SongEcho \cite{lisongecho} & 7.41 & 5.43 & 6.56 & 7.32 \\
ACE-Step-1.5 \cite{gong2026ace} & 8.23 & 5.71 & 7.62 & 7.82 \\
\ours{} & 8.31 & 6.27 & 7.66 & 7.91 \\
\bottomrule
\end{tabular*}
\end{minipage}
\par\addvspace{8pt}
\begin{minipage}{\linewidth}
\setlength{\parskip}{0pt}
\captionsetup{type=table}
\caption*{Table~\ref{tab:partial_outpainting} (continued).}
\fontsize{9}{11}\selectfont
\setlength{\tabcolsep}{3pt}
\renewcommand{\arraystretch}{1.08}
\noindent\textbf{SongEval aesthetics ($\uparrow$) \cite{yao2025songeval}}\par\smallskip
\begin{tabular*}{\textwidth}{@{\extracolsep{\fill}}lccccc@{}}
\toprule
Method & Coh. & Mem. & Nat. & Cla. & Mus. \\
\midrule
SongEcho \cite{lisongecho} & 2.87 & 2.76 & 2.63 & 2.58 & 2.67 \\
ACE-Step-1.5 \cite{gong2026ace} & 4.17 & 4.13 & 3.89 & 4.02 & 4.01 \\
\ours{} & 4.54 & 4.54 & 4.29 & 4.39 & 4.43 \\
\bottomrule
\end{tabular*}
\end{minipage}
\par\addvspace{8pt}
\begin{minipage}{\linewidth}
\setlength{\parskip}{0pt}
\captionsetup{type=table}
\caption*{Table~\ref{tab:partial_outpainting} (continued).}
\fontsize{9}{11}\selectfont
\setlength{\tabcolsep}{3pt}
\renewcommand{\arraystretch}{1.08}
\noindent\textbf{Lyrics, style, and cover identification}\par\smallskip
\begin{tabular*}{\textwidth}{@{\extracolsep{\fill}}lccccc@{}}
\toprule
Method & PER ($\downarrow$) & MuQ-Sim ($\uparrow$) & MuQ-Dist ($\uparrow$) & CSI-R@1 ($\uparrow$) & CSI-Rank ($\downarrow$) \\
\midrule
SongEcho \cite{lisongecho} & 0.64 & 0.20 & 0.72 & 0.25 & 7.61 \\
ACE-Step-1.5 \cite{gong2026ace} & 0.40 & 0.36 & 0.53 & 0.38 & 6.89 \\
\ours{} & 0.12 & 0.42 & 0.51 & 0.55 & 3.36 \\
\bottomrule
\end{tabular*}
\end{minipage}
\par\endgroup\addvspace{8pt}

\par\addvspace{8pt}
\begingroup
\captionsetup{type=table}
    \centering
    \begin{minipage}{\linewidth}
\setlength{\parskip}{0pt}
\caption{Partial-reference cover, \emph{inpainting} configuration.
        Same metrics as Table~\ref{tab:obj_main}.}
    \label{tab:partial_inpainting}
\fontsize{9}{11}\selectfont
\setlength{\tabcolsep}{3pt}
\renewcommand{\arraystretch}{1.08}
\noindent\textbf{AudioBox quality ($\uparrow$) \cite{tjandra2025meta_audiobox}}\par\smallskip
\begin{tabular*}{\textwidth}{@{\extracolsep{\fill}}lcccc@{}}
\toprule
Method & PQ & PC & CE & CU \\
\midrule
SongEcho \cite{lisongecho} & 7.83 & 5.90 & 7.20 & 7.56 \\
ACE-Step-1.5 \cite{gong2026ace} & 8.30 & 6.16 & 7.70 & 7.87 \\
\ours{} & 8.34 & 6.33 & 7.70 & 7.93 \\
\bottomrule
\end{tabular*}
\end{minipage}
\par\addvspace{8pt}
\begin{minipage}{\linewidth}
\setlength{\parskip}{0pt}
\captionsetup{type=table}
\caption*{Table~\ref{tab:partial_inpainting} (continued).}
\fontsize{9}{11}\selectfont
\setlength{\tabcolsep}{3pt}
\renewcommand{\arraystretch}{1.08}
\noindent\textbf{SongEval aesthetics ($\uparrow$) \cite{yao2025songeval}}\par\smallskip
\begin{tabular*}{\textwidth}{@{\extracolsep{\fill}}lccccc@{}}
\toprule
Method & Coh. & Mem. & Nat. & Cla. & Mus. \\
\midrule
SongEcho \cite{lisongecho} & 3.50 & 3.37 & 3.17 & 3.27 & 3.24 \\
ACE-Step-1.5 \cite{gong2026ace} & 4.31 & 4.30 & 4.06 & 4.16 & 4.19 \\
\ours{} & 4.58 & 4.58 & 4.33 & 4.43 & 4.46 \\
\bottomrule
\end{tabular*}
\end{minipage}
\par\addvspace{8pt}
\begin{minipage}{\linewidth}
\setlength{\parskip}{0pt}
\captionsetup{type=table}
\caption*{Table~\ref{tab:partial_inpainting} (continued).}
\fontsize{9}{11}\selectfont
\setlength{\tabcolsep}{3pt}
\renewcommand{\arraystretch}{1.08}
\noindent\textbf{Lyrics, style, and cover identification}\par\smallskip
\begin{tabular*}{\textwidth}{@{\extracolsep{\fill}}lccccc@{}}
\toprule
Method & PER ($\downarrow$) & MuQ-Sim ($\uparrow$) & MuQ-Dist ($\uparrow$) & CSI-R@1 ($\uparrow$) & CSI-Rank ($\downarrow$) \\
\midrule
SongEcho \cite{lisongecho} & 0.38 & 0.35 & 0.46 & 0.73 & 2.62 \\
ACE-Step-1.5 \cite{gong2026ace} & 0.32 & 0.32 & 0.46 & 0.82 & 1.68 \\
\ours{} & 0.13 & 0.41 & 0.49 & 0.93 & 1.28 \\
\bottomrule
\end{tabular*}
\end{minipage}
\par\endgroup\addvspace{8pt}

\FloatBarrier
\section{Subjective Evaluation: Listening Study}
\label{app:subjective_protocol}
\label{app:subjective_stats}

\begin{figure}[p]
    \centering
    \includegraphics[width=0.60\textwidth]{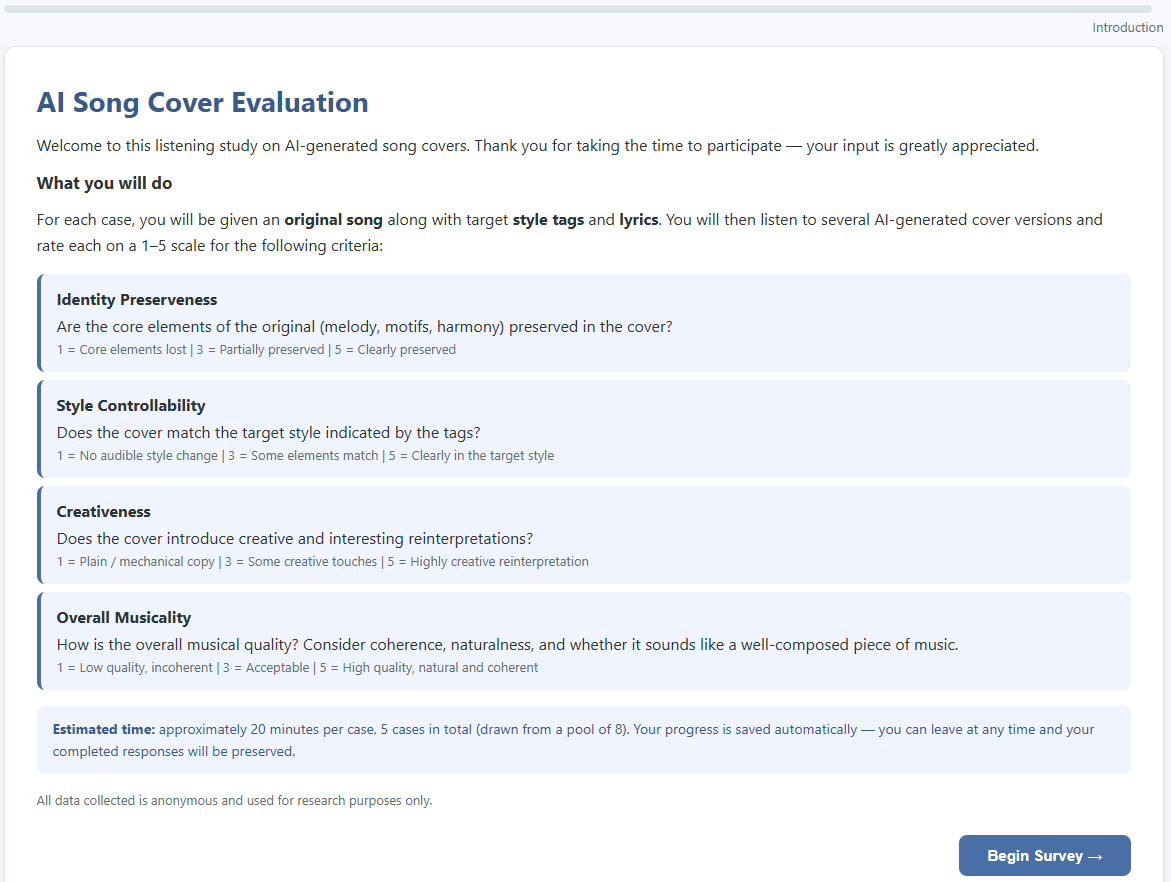}\\[3pt]
    \includegraphics[width=0.60\textwidth]{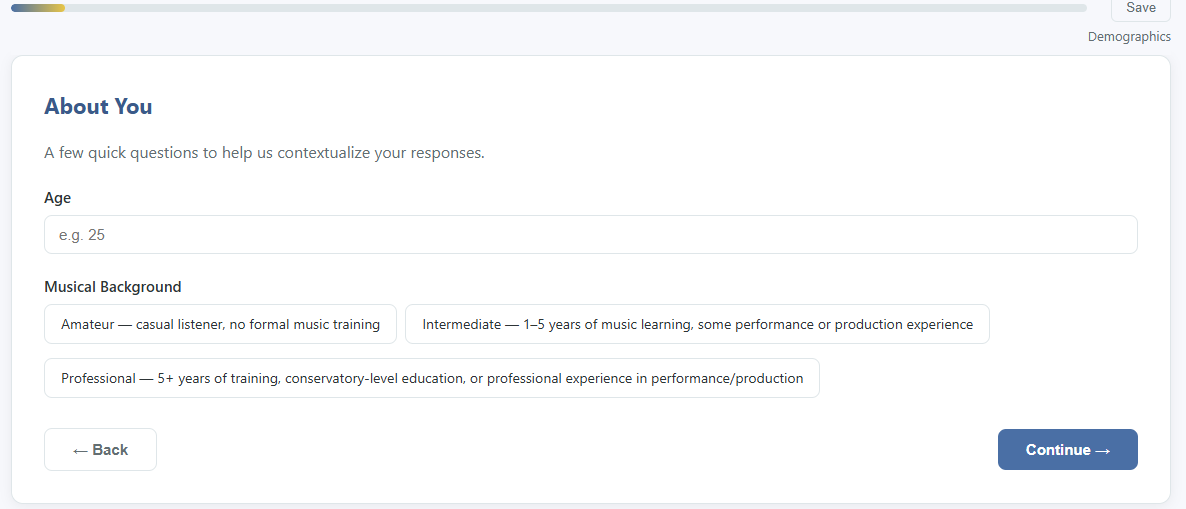}\\[3pt]
    \includegraphics[width=0.60\textwidth]{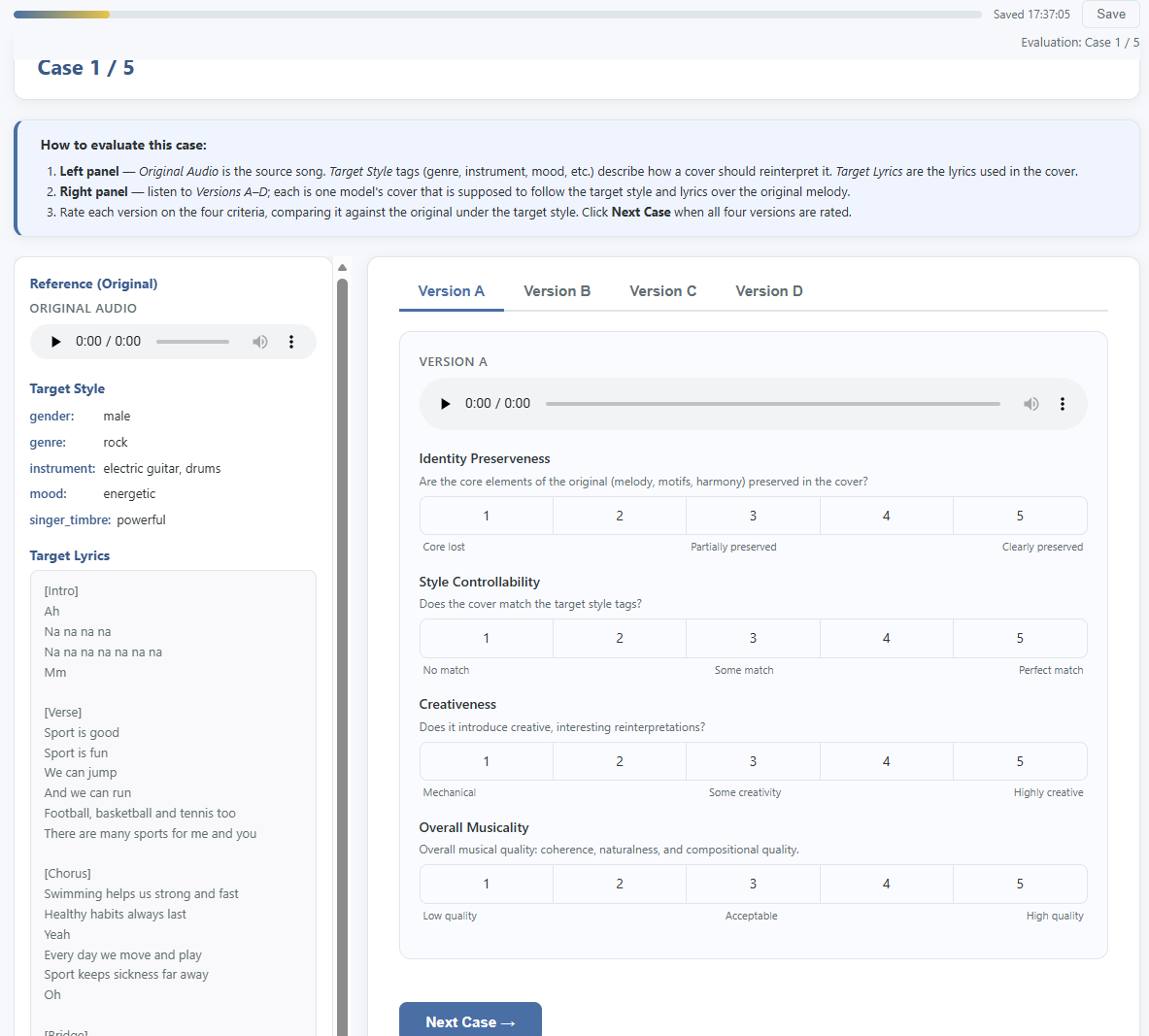}
    \caption{Listening-study questionnaire interface. Top: welcome page. Middle:
        demographics page. Bottom: per-case rating page, listing the four metrics
        together with the textual anchors for scores 1--5.}
    \label{fig:survey_screens}
\end{figure}

\paragraph{Protocol.}
On each rating page the participant is presented with the reference music, the
target style tag, and the target lyrics, and is asked to score the cover on four
5-point Likert scales: \emph{Identity Preserveness}, \emph{Style Controllability},
\emph{Creativeness}, and \emph{Overall Musicality}. The verbatim wording of each
metric and the anchor description for each of the scores 1--5 are shown on the
case page in Figure~\ref{fig:survey_screens}. Before the rating phase
the participant reports age and self-assessed musical background (amateur,
intermediate, or professional). Participants could stop at any point, with
completed responses auto-saved, so as to mitigate listening fatigue.
All participants in both the listening study and the expert interview were
unpaid volunteers and gave informed consent on the welcome page before
starting the questionnaire.

\paragraph{Participants.}
We received $34$ saved submissions. After excluding $12$ submissions with no
complete case and dropping incomplete case blocks, the analysis retained $22$
participants self-reporting their musical background as amateur~($11$),
intermediate~($6$), or professional~($5$). Among participants with valid age
responses ($n{=}18$), age was $25.5\pm 4.1$ (range $21$--$33$). Each retained
participant completed $4.32$ cases on average, yielding $95$ complete
(participant, case) blocks and $1{,}520$ ratings in total. A complete block is a
participant--case tuple in which all four systems were rated on all four
metrics. Table~\ref{tab:subj_background} breaks the average rating down by
self-reported musical background, pooled across systems.

\begin{table}[!htbp]
    \centering
    \caption{Mean rating per metric by self-reported musical background, pooled
        across systems, with cluster-bootstrap $95\%$ CI in brackets (resampling
        at the participant level, $1{,}000$ replicates). $n$ is the number of
        (participant, case, system) ratings contributing to each row, per metric.}
    \label{tab:subj_background}
    \setlength{\tabcolsep}{4pt}
    \resizebox{\textwidth}{!}{%
        \begin{tabular}{lccccc}
            \toprule
            Background    & $n$   & Identity Preserveness & Style Controllability & Creativeness          & Overall Musicality    \\
            \midrule
            Amateur       & $200$ & $3.54\,[3.27,\,3.79]$ & $3.38\,[3.12,\,3.62]$ & $3.21\,[3.00,\,3.40]$ & $3.50\,[3.35,\,3.67]$ \\
            Intermediate~ & $100$ & $3.25\,[2.54,\,3.80]$ & $3.08\,[2.37,\,3.70]$ & $3.16\,[2.69,\,3.58]$ & $3.26\,[2.82,\,3.64]$ \\
            Professional  & $~80$ & $3.26\,[2.82,\,3.88]$ & $3.36\,[2.67,\,4.05]$ & $3.17\,[2.67,\,3.51]$ & $2.89\,[2.28,\,3.34]$ \\
            \bottomrule
        \end{tabular}%
    }
\end{table}

\paragraph{Omnibus test across systems.}
Because every (participant, case) block receives one rating from each of the
four systems on each metric, the rating data is within-subject and paired. We
therefore use the complete participant--case tuple as the paired analysis block:
for each metric, the $k{=}4$ system ratings are ranked within each of the
$n_{\text{blocks}}{=}95$ blocks and tested for system-level rank differences
with a Friedman rank test. The $95$ complete blocks come from $112$ evaluated
cases after dropping $17$ incomplete cases, and each complete block contributes
$4{\times}4{=}16$ ratings.
Table~\ref{tab:subj_omnibus} reports the test statistic, $p$-value, and
Kendall's coefficient of concordance $W$ as effect size. Significant
system-level differences are observed on every one of the four metrics, with
$W$ ranging from $0.19$ (Creativeness) to $0.31$ (Overall Musicality),
motivating the pairwise comparisons below. We report these $p$-values as paired
block-level evidence of within-block system differences, rather than as
evidence from $95$ fully independent subject-level observations.

\begin{table}[!htbp]
    \centering
    \caption{Friedman omnibus test across the four systems, per metric.
        $n_{\text{blocks}}$ is the number of (participant, case) blocks within
        which the $k{=}4$ system ratings are ranked. Kendall's $W$ is the
        coefficient-of-concordance effect size.}
    \label{tab:subj_omnibus}
    \setlength{\tabcolsep}{8pt}
    \begin{tabular}{lcccc}
        \toprule
        Metric                & $n_{\text{blocks}}$ & $\chi^2$ & $p$-value             & Kendall's $W$ \\
        \midrule
        Identity Preserveness & $95$                & $57.19$  & $2.3{\times}10^{-12}$ & $0.20$        \\
        Style Controllability & $95$                & $55.69$  & $4.9{\times}10^{-12}$ & $0.20$        \\
        Creativeness          & $95$                & $52.90$  & $1.9{\times}10^{-11}$ & $0.19$        \\
        Overall Musicality    & $95$                & $88.08$  & $5.7{\times}10^{-19}$ & $0.31$        \\
        \bottomrule
    \end{tabular}
\end{table}

\paragraph{Pairwise post-hoc tests.}
Conditional on the omnibus rejecting on every metric, we run pairwise
Wilcoxon signed-rank tests on the within-block (participant, case) score
differences for each of the $\binom{4}{2}{=}6$ system pairs, and apply a
Bonferroni correction over the $6$ comparisons within each metric.
Table~\ref{tab:subj_posthoc} reports the corrected $p$-values; bold entries
are significant at the $0.05$ level. Three observations follow.
\emph{(i)}~On Style Controllability, Creativeness, and Overall Musicality,
\ours{} shows no significant difference from Suno-v5.5
($p_{\text{bonf}}{=}1.0$ on all three) while significantly outperforming both
ACE-Step-1.5 ($p_{\text{bonf}}{\leq}0.017$) and SongEcho
($p_{\text{bonf}}{<}10^{-4}$ on all three).
\emph{(ii)}~On Identity Preserveness, \ours{} significantly outperforms
SongEcho ($p_{\text{bonf}}{=}0.0001$) and shows no significant difference from
ACE-Step-1.5 ($p_{\text{bonf}}{=}1.0$), while Suno-v5.5 remains significantly
higher than \ours{} ($p_{\text{bonf}}{=}0.006$).
\emph{(iii)}~SongEcho is significantly lower than \ours{} and Suno-v5.5 on all
four metrics; compared with ACE-Step-1.5, SongEcho is lower on Identity
Preserveness, Creativeness, and Overall Musicality, but the Style
Controllability difference is not significant.

\begin{table}[!htbp]
    \centering
    \caption{Bonferroni-corrected $p$-values from pairwise Wilcoxon
        signed-rank tests across the four systems, per metric. Each test pairs
        the two systems on the within-block (participant, case) score
        differences over the $n_{\text{blocks}}{=}95$ complete blocks; the
        Bonferroni correction is applied over the $6$ pairs within each
        metric. Bold marks $p<0.05$.}
    \label{tab:subj_posthoc}
    \setlength{\tabcolsep}{6pt}
    \begin{tabular}{lcccc}
        \toprule
        Pair                        & Identity            & Style               & Creativeness        & Overall             \\
        \midrule
        ACE-Step-1.5 vs.\ \ours{}   & $1.0000$            & $\mathbf{<10^{-4}}$ & $\mathbf{0.0004}$   & $\mathbf{0.0161}$   \\
        ACE-Step-1.5 vs.\ SongEcho  & $\mathbf{<10^{-4}}$ & $1.0000$            & $\mathbf{0.0169}$   & $\mathbf{<10^{-4}}$ \\
        ACE-Step-1.5 vs.\ Suno-v5.5 & $\mathbf{0.0475}$   & $\mathbf{<10^{-4}}$ & $\mathbf{0.0007}$   & $\mathbf{<10^{-4}}$ \\
        \ours{} vs.\ SongEcho       & $\mathbf{0.0001}$   & $\mathbf{<10^{-4}}$ & $\mathbf{<10^{-4}}$ & $\mathbf{<10^{-4}}$ \\
        \ours{} vs.\ Suno-v5.5      & $\mathbf{0.0060}$   & $1.0000$            & $1.0000$            & $1.0000$            \\
        SongEcho vs.\ Suno-v5.5     & $\mathbf{<10^{-4}}$ & $\mathbf{<10^{-4}}$ & $\mathbf{<10^{-4}}$ & $\mathbf{<10^{-4}}$ \\
        \bottomrule
    \end{tabular}
\end{table}

\FloatBarrier
\section{Subjective Evaluation: Expert Interviews}
\label{app:expert_interviews}

\paragraph{Protocol.}
The expert interview is designed to explain the Likert results without letting
the evaluator choose criteria after hearing the generated audio. In each
session, we first sample a target genre from a catalogue of common
popular-music styles. Before any audio is played, the expert is asked to list
the most salient musical attributes of that genre and briefly explain how each
attribute should be judged. Only after these attributes are fixed do we present
the blinded covers from the four systems. The expert then scores and comments
on each version using the pre-elicited attributes as the evaluation criteria.
Finally, we feed the organized expert record to a language model and ask it to
produce a concise, neutral summary. The respondent identifier and date are
anonymized in the record below.

\paragraph{R\&B case study.}
The sampled genre in the recorded session is R\&B. The four blinded versions
were later decoded as Version~A = ACE-Step-1.5, Version~B = SongEcho,
Version~C = \ours{}, and Version~D = Suno-v5.5. The tables and notes below are
translated from the original Chinese notes and lightly edited for readability
while preserving the expert's recorded ratings and judgments.

\begin{table}[!htbp]
    \centering
    \caption{Primary R\&B attributes elicited from the expert before any
        generated audio was played.}
    \label{tab:expert_rnb_attributes}
    \setlength{\tabcolsep}{5pt}
    \begin{tabular}{>{\raggedright\arraybackslash}p{0.25\textwidth}>{\raggedright\arraybackslash}p{0.60\textwidth}}
        \toprule
        Attribute                               & Expert definition / judging criterion \\
        \midrule
        Rhythm and groove                       &
        The most salient property of rhythm-and-blues: drums and bass jointly
        form the groove, and faithful R\&B should exhibit typical rhythmic
        patterns rather than a generic pop or electronic feel.                          \\
        Vocal melody, repetition, and variation &
        In R\&B,
        melodic development often happens through modular repetition and varied
        repetition over a cyclic rhythmic/harmonic frame; phrase-end ornaments,
        melismas, and vocal timbre are important cues.                                  \\
        \bottomrule
    \end{tabular}
\end{table}

\begin{table}[!htbp]
\centering
\caption{Organized expert record for the R\&B case study. Each attribute
    follows the pre-listening criteria in
    Table~\ref{tab:expert_rnb_attributes}. The table focuses on the two
    primary attributes for which the expert gave complete scores.}
\label{tab:expert_rnb_record}
\begingroup
\small
\setlength{\tabcolsep}{3pt}
\renewcommand{\arraystretch}{1.15}
\begin{tabular}{>{\raggedright\arraybackslash}p{0.18\textwidth}>{\raggedright\arraybackslash}p{0.25\textwidth}>{\raggedright\arraybackslash}p{0.23\textwidth}>{\raggedright\arraybackslash}p{0.25\textwidth}}
    \toprule
    Version / system                                                      & Rhythm and groove & Vocal melody / timbre & Overall comment \\
    \midrule
    A / ACE-Step-1.5                                                      &
    $2$: Electric drums, 808, and synthetic bass make the version sound
    closer to electronic pop.                                             &
    $2$: The vocal line is closer to a normal pop-song melody than to
    R\&B melodic development.                                             &
    The first two attributes deviate substantially from typical R\&B; the
    version leans toward electronic pop rather than canonical R\&B.                                                                     \\
    \midrule
    B / SongEcho                                                          &
    $3$: The drum timbre and groove contain pop-R\&B characteristics.     &
    $3$: The vocal development is closer to R\&B: it uses a four-phrase
    loop rather than the linear emotional buildup typical of pop ballads. &
    Overall, this version is relatively better than Version~A.                                                                          \\
    \midrule
    C / \ours{}                                                           &
    $4$: The expert rated the rhythm and groove higher than Versions~A and
    B.                                                                    &
    $4$: The expert rated the vocal melodic treatment higher than
    Versions~A and B; the verse also has a vocal timbre that sounds very
    R\&B-like.                                                            &
    The expert judged it close to Version~B in broad style but with a more
    R\&B-like singer timbre; it is still not the most canonical R\&B
    rendition.                                                                                                                          \\
    \midrule
    D / Suno-v5.5                                                         &
    $4.5$: The groove is slightly better than the preceding versions.     &
    $4.5$: The vocal line is much more natural and close to pop R\&B,
    especially in phrase endings, melismas, and timbre.                   &
    The expert also noted that the auxiliary harmonic and
    instrumentation-related attributes were relatively well matched.                                                                    \\
    \bottomrule
\end{tabular}
\endgroup
\end{table}

The expert also listed harmonic looping and instrumentation as auxiliary
references, but described them as less diagnostic than rhythm/groove and vocal
melodic treatment. We therefore report only the two primary criteria in
Table~\ref{tab:expert_rnb_record}. In the original notes, the auxiliary
comments are sparse: ACE-Step-1.5 receives a harmony score of 3 without an
additional written rationale, and Suno-v5.5 is described in the overall comment
as relatively well matched on the harmonic and instrumentation-related cues.

\Needspace{10\baselineskip}
\paragraph{Prompt used for the language-model summary.}
For step~(iv), we used the following prompt template to convert the organized
expert record into a concise verdict. To avoid first-person or system-ownership
bias, the prompt asks the language model to reason over the blinded version
labels only. The version labels were mapped back to system names after the
summary was produced.

\begin{center}
\begingroup
\setlength{\fboxsep}{6pt}
\setlength{\fboxrule}{0.4pt}
\fbox{%
    \begin{minipage}{0.92\textwidth}
        \small
        \textbf{Input:} A structured expert-interview record for an AI song-cover
        evaluation. The record contains: (1) the sampled target genre; (2) the musical
        attributes elicited from the expert before any generated audio was played; and
        (3) blinded versions A--D with per-attribute scores and expert comments.

        \textbf{Task:} Produce a concise 4--6 sentence neutral research summary
        contrasting Versions A--D. Base the verdict only on the expert's stated
        attributes, scores, and comments. Mention the genre-specific criteria that
        drive the ranking. Do not add claims that are not supported by the expert
        record. Do not use first-person labels such as ``our model'' or infer that any
        version should be preferred from its label.

        \textbf{Desired output:} A short neutral paragraph contrasting Version~A,
        Version~B, Version~C, and Version~D.
    \end{minipage}%
}
\endgroup
\end{center}

\FloatBarrier
\FloatBarrier
\section{Stage~3 Cross-Cover Data Pipeline}
\label{app:stage3_pipeline}

We prompt an LLM to generate $1{,}250$ lyric sets paired with HeartMuLa-format style tags (also LLM-generated), and feed them to music generation providers to synthesize $1{,}250$ source songs. For each source song, we then feed its audio, its lyrics, and seven additional, pairwise-distinct LLM-generated style tags to music generation providers to produce seven cover versions; 
Because each music generation provider returns two songs, the procedure above is effectively doubled, yielding a cross-cover dataset of $20{,}000$ paired covers.

For training, we form pairs by sampling two distinct covers of the same source song. Our network is sensitive to global key, so a cover whose key has shifted during cover generation is unusable for training. We therefore extract lead sheets from both covers in each candidate pair, run the key detection of \cite{ou2024unifying}, and discard pairs whose keys disagree. From only $20{,}000$ songs this yields roughly $130{,}000$ trainable pairs, substantially improving data utilization and strengthening the style controllability of the model.

\FloatBarrier
\section{Generalization Ability on Human Recordings}

\ours{} is trained on synthetic data. To evaluate whether our approach generalizes beyond training distribution, we additionally conduct an experiment on real human recordings.

We manually choose 8 songs with 4 Chinese songs and 4 English songs. We use SongEcho \cite{lisongecho} and ACE-Step 1.5 \cite{gong2026ace} as baselines. Suno-v5.5 is excluded since generating cover versions from human recordings is not supported.

The results are shown in Table \ref{tab:obj_human}. \ours{} consistently outperforms baselines on most metrics, which suggests that our approach can generalize to unseen human-composed tracks despite primarily trained on synthetic data. We note that this evaluation is small in scale, and leave a more comprehensive study on real-world data to future work.
\begin{table}[t]
    \centering
    \caption{Objective evaluation on the full song cover task
        (Task~1) on human-recordings. $\uparrow$ / $\downarrow$
        indicate higher / lower is better. \textbf{Bold} marks the best.}
    \label{tab:obj_human}
    \resizebox{\textwidth}{!}{
        \begin{tabular}{lccccccccccccccc}
            \toprule
                                            & \multicolumn{4}{c}{AudioBox \cite{tjandra2025meta_audiobox} ($\uparrow$)} & \multicolumn{5}{c}{SongEval \cite{yao2025songeval} ($\uparrow$)} & Lyrics        & \multicolumn{2}{c}{Style} & \multicolumn{2}{c}{CSI judge}                                                                                                                                                                                      \\
            \cmidrule(lr){2-5} \cmidrule(lr){6-10} \cmidrule(lr){11-11} \cmidrule(lr){12-13} \cmidrule(lr){14-15}
            Method                          & PQ                                                                        & PC                                                               & CE            & CU                        & Coh.                          & Mem.          & Nat.          & Cla.          & Mus.          & PER ($\downarrow$) & MuQ-Sim ($\uparrow$) & MuQ-Dist ($\uparrow$) & CSI-R@1 ($\uparrow$) & CSI-Rank ($\downarrow$) \\
            \midrule
            SongEcho \cite{lisongecho}      & 7.66                                                                      & \textbf{6.26}                                                             & 7.06          & 7.32                      & 3.47                          & 3.23         & 3.13          & 3.19          & 3.27          & 0.42               & 0.32                 & 0.43         & 0.97                 & 1.03                    \\
            ACE-Step-1.5 \cite{gong2026ace} & 8.27                                                                      & 6.08                                                             & 7.67          & 7.86                      & 4.37                          & 4.36          & 4.15          & 4.25          & 4.25          & 0.25               & 0.31                 & 0.31                  & 0.98                 & 1.08                    \\
            \midrule
            \ours{}                         & \textbf{8.34}                                                             & 6.24                                                   & \textbf{7.73} & \textbf{7.92}             & \textbf{4.52}                 & \textbf{4.51} & \textbf{4.27} & \textbf{4.37} & \textbf{4.40} & \textbf{0.12}      & \textbf{0.40}        & \textbf{0.45}                 & \textbf{0.98}        & \textbf{1.02}           \\
            \bottomrule
        \end{tabular}
    }
\end{table}

\section{Preservation of Basic Text-to-Music Ability}

\ours{} incorporates classifier-free guidance for lead-sheet conditioning. Therefore, to investigate whether the original text-to-music (T2M) capability is preserved, we replace the input lead-sheet condition with a null condition (denoted as \ours{}-nc). Without the lead-sheet condition, \ours{}-nc naturally falls back to a pure text-to-music generation setting. We evaluate \ours{}-nc and the original HeartMuLa model on the same test set. Results are shown in Table \ref{tab:obj_nc}.

Overall, \ours{}-nc achieves comparable performance to the original HeartMuLa model across most metrics, indicating that the LoRA fine-tuning does not introduce significant catastrophic forgetting of the original T2M capability. Compared with HeartMuLa, \ours{}-nc achieves slightly better performance on three out of four AudioBox metrics (PQ, CE, and CU), while showing moderate degradation on some SongEval dimensions (Coh., Mem., Nat., Cla., and Mus.). Meanwhile, the lyric and style-following metrics (PER and MuQ-Sim) are improved after fine-tuning.

We hypothesize that this improvement may be related to our training objective in Stage 3. During this stage, the model learns to generate cover versions conditioned on the lyrics and style tags of the target cover song, while the lead-sheet condition is extracted from the original song. Since the textual conditions and lead-sheet condition correspond to different versions of the song, the model is encouraged to better disentangle and utilize textual style/lyric information rather than relying solely on the musical condition. This may improve its ability to follow textual conditions even in the absence of lead-sheet input.
\begin{table}[t]
    \centering
    \caption{Comparison between \ours{}-nc and HeartMuLa \cite{yang2026heartmula}. $\uparrow$ /
        $\downarrow$ indicate higher / lower is better. \textbf{Bold} marks
        the better result in each column.}
    \label{tab:obj_nc}
    \resizebox{\textwidth}{!}{
        \begin{tabular}{lcccccccccccc}
            \toprule
             & \multicolumn{4}{c}{AudioBox \cite{tjandra2025meta_audiobox} ($\uparrow$)}
             & \multicolumn{5}{c}{SongEval \cite{yao2025songeval} ($\uparrow$)}
             & Lyrics
             & \multicolumn{2}{c}{Style} \\
            \cmidrule(lr){2-5} \cmidrule(lr){6-10} \cmidrule(lr){11-11} \cmidrule(lr){12-13}
            Method & PQ & PC & CE & CU & Coh. & Mem. & Nat. & Cla. & Mus.
                   & PER ($\downarrow$)
                   & MuQ-Sim ($\uparrow$) & MuQ-Dist ($\uparrow$) \\
            \midrule
            HeartMuLa \cite{yang2026heartmula}
                & 8.23 & \textbf{6.32} & 7.61 & 7.89
                & \textbf{4.68} & \textbf{4.69} & \textbf{4.51} & \textbf{4.58} & \textbf{4.60}
                & 0.15
                & 0.27 & \textbf{0.59} \\
            \midrule
            \ours{}-nc
                & \textbf{8.30} & 6.18 & \textbf{7.68} & \textbf{7.92}
                & 4.60 & 4.59 & 4.34 & 4.43 & 4.47
                & \textbf{0.11}
                & \textbf{0.41} & \textbf{0.59} \\
            \bottomrule
        \end{tabular}
    }
\end{table}
\section{Robustness to Noisy Lead-sheets}

\label{sec:robustness}

We investigate two common types of lead-sheet extraction errors, namely
missing notes and inaccurate notes, and provide a brief robustness analysis.
To simulate missing notes, each note in the extracted lead sheet is randomly
removed with probability $p_1$. To simulate inaccurate notes, we first sample
a pitch shift (in semitones) uniformly from $[-3, 3]$ for each note, and then
shift the note by this amount with probability $p_2$. We evaluate \ours{}
on Task~1 under different $(p_1, p_2)$ settings, as reported in
Table~\ref{tab:robustness}.

\begin{table}[t]
    \centering
    \caption{Robustness of \ours{} to lead-sheet extraction errors on Task~1.
        $p_1$ and $p_2$ denote the probabilities of note removal and pitch
        shifting, respectively. $\uparrow$ / $\downarrow$ indicate higher /
        lower is better.}
    \label{tab:robustness}
    \resizebox{\textwidth}{!}{
        \begin{tabular}{cccccccccccccccc}
            \toprule
            \multicolumn{2}{c}{Corruption}
             & \multicolumn{4}{c}{AudioBox ($\uparrow$)}
             & \multicolumn{5}{c}{SongEval ($\uparrow$)}
             & Lyrics
             & \multicolumn{2}{c}{Style}
             & \multicolumn{2}{c}{CSI judge} \\
            \cmidrule(lr){1-2} \cmidrule(lr){3-6} \cmidrule(lr){7-11} \cmidrule(lr){12-12} \cmidrule(lr){13-14} \cmidrule(lr){15-16}
            $p_1$ & $p_2$ & PQ & PC & CE & CU & Coh. & Mem. & Nat. & Cla. & Mus.
                  & PER ($\downarrow$)
                  & MuQ-Sim ($\uparrow$) & MuQ-Dist ($\uparrow$)
                  & CSI-R@1 ($\uparrow$) & CSI-Rank ($\downarrow$) \\
            \midrule
            0      & 0      & 8.34 & 6.34 & 7.70 & 7.93 & 4.59 & 4.59 & 4.34 & 4.44 & 4.47 & 0.13 & 0.41 & 0.48 & 0.97 & 1.05 \\
            \midrule
            0      & 5\%    & 8.29 & 6.26 & 7.67 & 7.90 & 4.54 & 4.52 & 4.25 & 4.35 & 4.40 & 0.13 & 0.40 & 0.56 & 0.63 & 2.85 \\
            0      & 10\%   & 8.26 & 6.22 & 7.61 & 7.87 & 4.48 & 4.46 & 4.18 & 4.29 & 4.34 & 0.14 & 0.40 & 0.57 & 0.42 & 5.10 \\
            5\%    & 0      & 8.26 & 6.20 & 7.61 & 7.89 & 4.51 & 4.50 & 4.23 & 4.32 & 4.37 & 0.13 & 0.39 & 0.56 & 0.62 & 2.91 \\
            10\%   & 0      & 8.29 & 6.26 & 7.70 & 7.91 & 4.57 & 4.56 & 4.28 & 4.39 & 4.44 & 0.11 & 0.38 & 0.57 & 0.44 & 3.69 \\
            10\%   & 10\%   & 8.30 & 6.23 & 7.70 & 7.92 & 4.52 & 4.51 & 4.23 & 4.33 & 4.37 & 0.12 & 0.38 & 0.58 & 0.28 & 5.93 \\
            \bottomrule
        \end{tabular}
    }
\end{table}

Overall, \ours{} is robust to moderate lead-sheet extraction errors in terms
of generation quality: both AudioBox and SongEval scores exhibit only limited
degradation across all corruption levels, and lyric accuracy remains stable.
In contrast, cover identification performance decreases as the corruption
level increases. This is expected, since a less accurate lead sheet provides
weaker guidance on the musical identity of the reference song.
Interestingly, MuQ-Dist increases under corrupted conditions. We hypothesize
that noisy conditions introduce additional deviations from the original song,
encouraging more diverse reinterpretations and thus increasing the
distributional distance between the generated cover and the reference.

\section{Lead-sheet Controllability}
\begin{table}[t]
    \centering
    \caption{Lead sheet metrics on the full-song cover task (Task~1).
        $\uparrow$ indicates higher is better. \textbf{Bold} marks the best
        and \underline{underline} the second-best result in each column.}
    \label{tab:obj_leadsheet}
    \begin{tabular}{lcc}
        \toprule
         & \multicolumn{2}{c}{Lead sheet} \\
        \cmidrule(lr){2-3}
        Method & MS ($\uparrow$) & CS ($\uparrow$) \\
        \midrule
        SongEcho \cite{lisongecho}        & \textbf{0.77}    & 0.74 \\
        ACE-Step-1.5 \cite{gong2026ace}   & 0.63             & \underline{0.86} \\
        Suno-v5.5 \cite{Suno_v5_5}        & 0.57             & 0.82 \\
        \midrule
        \ours{}                           & \underline{0.68} & \textbf{0.89} \\
        \bottomrule
    \end{tabular}
\end{table}

We additionally report lead sheet controllability for Task 1. We consider melody similarity (MS) and chord similarity (CS). Specifically, we extract the frame-level melody and chord sequences from the generated song, represent them as pitch/chroma vectors, and compute the frame-wise cosine similarity with the corresponding sequences of the input lead sheet, averaged over all frames. For models whose outputs are not frame-aligned with the input (\ours{} and Suno-v5.5), we first align the two sequences with Dynamic Time Warping (DTW) and then compute the similarity along the optimal alignment path. See Table \ref{tab:obj_leadsheet} for results.

For lead-sheet control, \ours{} follows the input chord progression most closely of all methods. SongEcho attains higher melody similarity because it conditions on frame-wise F0 from the source vocals, which closely matches the frame-level metric, whereas our tempo-normalized melody condition allows rhythmic flexibility at the cost of a slight alignment error. Among methods without frame-level alignment, \ours{} still follows the melody more closely than Suno-v5.5.

\begingroup
\setlength{\bibsep}{2pt}
\setlength{\parskip}{0pt}
\bibliographystyle{unsrtnat}
\bibliography{refs}
\endgroup
\FloatBarrier
\section*{Contributors}
Lynn Yi, Zihan Xiong, Junjie Cao, Xiaolong Weng, Yingzhe Ma, Gus Xia,
Jinliang Liu, Yuxin Xie, Xianwei Zhuang, Yuguo Yin, Qiao Jin, Gongxi Zhu,
Jiayu Wang, Junjie Liang, Qin Yue, Ziyu Wang, Dading Chong, Meng Cao,
Jiahuan Zhou, Dongchao Yang.

\end{document}